\documentclass[acmsmall,screen]{acmart}

\setcopyright{rightsretained}
\acmPrice{}
\acmDOI{10.1145/3408995}
\acmYear{2020}
\copyrightyear{2020}
\acmSubmissionID{icfp20main-p65-p}
\acmJournal{PACMPL}
\acmVolume{4}
\acmNumber{ICFP}
\acmArticle{113}
\acmMonth{8}

\usepackage{booktabs}   
\usepackage{subcaption} 

\usepackage{wrapfig}
\usepackage{url}
\usepackage{enumitem}
\usepackage{hyperref}
\usepackage{listings}
\usepackage{xcolor}
\usepackage{xspace}
\usepackage{cleveref}
\usepackage{multirow}
\usepackage{microtype}

\newcommand{\kc}[1]{{\color{red} {\it [KC says: #1]}}}

\newcommand{\cf}[1]{\texttt{#1}}

\definecolor{Bittersweet}{rgb}{1.0, 0.44, 0.37}
\definecolor{MidnightBlue}{rgb}{0.0, 0.2, 0.4}
\definecolor{BrightBlue}{rgb}{0.0, 0.2, 0.7}
\definecolor{byzantine}{rgb}{0.74, 0.2, 0.64}
\definecolor{caribbeangreen}{rgb}{0.0, 0.8, 0.6}

\lstnewenvironment{code}[1][]
    { 
      \lstset{
				basicstyle=\ttfamily\footnotesize,
				#1}%
      \csname lst@setfirstlabel\endcsname}
    { 
      \csname lst@savefirstlabel\endcsname}

\lstMakeShortInline[columns=fullflexible]|

\newcommand{\ml}[1]{\lstinline[language=caml]{#1}}

\def\techreport{1}

\begin{document}

\title{Retrofitting Parallelism onto OCaml}


\author{KC Sivaramakrishnan}
\affiliation{
  \institution{IIT Madras}            
	\country{India}
}
\email{kcsrk@cse.iitm.ac.in}          

\author{Stephen Dolan}
\affiliation{
  \institution{OCaml Labs}            
	\country{UK}
}
\email{stephen.dolan@cl.cam.ac.uk}

\author{Leo White}
\affiliation{
  \institution{Jane Street}            
	\country{UK}
}
\email{leo@lpw25.net}

\author{Sadiq Jaffer}
\affiliation{
  \institution{Opsian}                
	\country{UK}
}
\affiliation{
  \institution{OCaml Labs}            
	\country{UK}
}
\email{sadiq@toao.com}

\author{Tom Kelly}
\affiliation{
  \institution{OCaml Labs}            
	\country{UK}
}
\email{tom.kelly@cantab.net}

\author{Anmol Sahoo}
\affiliation{
  \institution{IIT Madras}            
	\country{India}
}
\email{anmol.sahoo25@gmail.com}          

\author{Sudha Parimala}
\affiliation{
  \institution{IIT Madras}            
	\country{India}
}
\email{sudharg247@gmail.com}          

\author{Atul Dhiman}
\affiliation{
  \institution{IIT Madras}            
	\country{India}
}
\email{atuldhiman94@gmail.com}          

\author{Anil Madhavapeddy}
\affiliation{
  \institution{University of Cambridge Computer Laboratory}            
	\country{UK}
}
\affiliation{
  \institution{OCaml Labs}            
	\country{UK}
}
\email{avsm2@cl.cam.ac.uk}          

\authorsaddresses{\em{The corresponding authors of this paper are KC Sivaramakrishnan (\url{kcsrk@cse.iitm.ac.in}) and Anil Madhavapeddy (\url{avsm2@cl.cam.ac.uk}).}}
\begin{abstract}
	OCaml is an industrial-strength, multi-paradigm programming language, widely
	used in industry and academia. OCaml is also one of the few modern managed
	system programming languages to lack support for shared memory parallel
	programming. This paper describes the design, a full-fledged implementation
	and evaluation of a mostly-concurrent garbage collector (GC) for the
	multicore extension of the OCaml programming language. Given that we propose
	to add parallelism to a widely used programming language with millions of
	lines of existing code, we face the challenge of maintaining backwards
	compatibility--not just in terms of the language features but also the
	performance of single-threaded code running with the new GC. To this end, the
	paper presents a series of novel techniques and demonstrates that the new GC
	strikes a balance between performance and feature backwards compatibility for
	sequential programs and scales admirably on modern multicore processors.
\end{abstract}


\maketitle
\renewcommand{\shortauthors}{Sivaramakrishnan et al.}

\section{Introduction}
\label{sec:intro}

Support for shared-memory parallelism is quite standard in managed system
programming languages nowadays. Languages and runtimes such as Go, Haskell,
Java, the .NET CLR platform and various parallel extensions of the ML
programming language~\cite{Westrick20,ueno16gc,Sivaramakrishnan14,Fluet10} all
support multiple parallel threads of execution. There has been extensive
research and development into efficient garbage collector (GC) designs for
these languages and runtimes. Our challenge is to retrofit parallelism to the
OCaml programming language, which has been in continuous use since 1996 in
large codebases, particularly in verification tools or mission-critical systems
components. Adding shared-memory parallelism to an existing language presents
an interesting set of challenges. As well as the difficulties of memory
management in a parallel setting, we must maintain as much backwards
compatibility as practicable. This includes not just compatibility of the
language semantics, but also of the performance profile, memory usage and C
bindings.

\subsection{Performance backwards compatibility}

The main challenge in adding support for shared-memory parallelism to OCaml is
the implementation of the multicore-capable GC. OCaml users have come to rely
on allocation and memory access being relatively cheap (for a managed language)
in order to write performant code.  Being a functional programming language,
OCaml code usually exhibits a high rate of allocation with most objects being
small and short-lived. Hence, OCaml uses a small minor heap with a generational
GC to collect the small, short-lived objects. Objects are allocated in the
minor heap by bumping the allocation pointer. Hence, the minor heap allocations
are very fast.

Many objects in OCaml are immutable. For immutable objects, the initialising
writes (the only ones) are done without barriers and reads require no barriers.
For mutable fields, the writes require a
(deletion/Yuasa/snapshot-at-the-beginning~\cite{Yuasa90}) barrier, and the
reads do not. Hence, reads are fast and updates are comparatively slower.

\if{0}
\kc{Much of what is said above sound like aphorisms. It would be useful to
produce actual numbers to justify these claims. I am thinking (1) a graph
showing object size distribution (x-axis size of objects in log scale, y-axis
cdf) (2) Average minor heap survival rate (3) allocation rates (which are
available from the PLDI paper) (4) distribution of mutable and immutable
objects (this would be harder to get). Now that we have sandmark, this should
be possible? }.
\fi

Objects that survive a minor collection are promoted to the major heap, which
is collected by an incremental, non-moving, mark-and-sweep collector, with an
optional compaction phase. This design minimizes pause times, and indeed, OCaml
is used to write latency sensitive applications such as network services
(MirageOS~\cite{Madhavapeddy13}) and user interfaces
(ReasonML~\cite{ReasonML}). Our aim with the multicore capable GC is to
preserve the fast performance characteristics of operations as they currently
exist on single-core OCaml, with some compromises for mutable objects to
deal with the possibility of racy access between multiple mutator and GC
threads. We strive to retain the low pause times of single-core OCaml.

An alternative way to achieve performance backwards compatibility for those
programs that do not utilise parallelism would be to have two separate
compilers and runtimes for the serial and parallel code. For example, GHC
Haskell comes with two separate runtimes, one which supports parallelism and
another which is purely single-threaded~\cite{Marlow09}. GHC provides a
compiler flag \cf{-threaded} to choose the runtime system to link to. In order
to avoid the maintenance burden of two separate runtimes, we chose to go for a
unified runtime.

\subsection{Feature backwards compatibility}

Given the millions of lines of OCaml code in production, it would serve us well
to ensure that the addition of parallelism to OCaml breaks as little code as
possible. OCaml is a type-safe language and the addition of parallelism should
not break type safety under data races. Dolan et al. present a memory model
for shared-memory parallel programs that gives strong guarantees (including
type safety) even in the presence of data races~\cite{Dolan18}. They describe
the instantiation of the memory model for the OCaml programming language and
demonstrate that the memory model has reasonable overheads. Our work inherits
this memory model.

Beyond type safety, OCaml has several features that closely interact with the
garbage collector. These include weak references, finalisers,
ephemerons~\cite{Hayes97}, and lazy values, whose semantics will have to be
preserved with the new GC so as to not break programs that use those features.
OCaml's C API exposes quite a lot of the internals of the memory
representation, and has an expressive API to access the heap
efficiently~\cite{OCamlCAPI}. While this allows users to write fast code, the
API also bakes in the invariants of the GC. For example, reading any OCaml
object, be it mutable or not, using the C API does not involve a read barrier,
and is compiled as a plain read of memory. A new GC scheme that adds a read
barrier only to reads of mutable fields will need to deprecate the old API or
suffer the risk of breaking code silently. Either way, the users will have to
modify their code to work correctly under the new GC. Given that the compiler
does not check incorrect uses of the C API, it is already difficult to write
correct and efficient FFI code. We would like to strike a balance between the
added complexity of the C API and the performance impact of the missed
opportunities.

\subsection{Requirements}
\label{sec:req}

In summary, we strive towards the following ideal goals in our parallel
extension of OCaml:

\begin{enumerate}[label=R\arabic*]
	\item A well behaved serial program does not break on the parallel extension.
		That is, a well-typed serial program remains well-typed in the parallel
		extension, and the semantics of such a program remains the same on the
		serial and parallel runtimes.
	\item The performance profile of a serial program on the parallel runtime
		remains the same as the serial runtime. That is, the program on the
		parallel runtime should run as fast as it does on serial runtime.
		Additionally, the GC pause times of the serial program on the parallel
		runtime remain the same as the serial runtime.
	\item The parallel programs should aim to minimize pause times, and then aim
		to run as fast as possible on the available cores. We order the sub goals
		this way since minimising pause times in the GC is much harder than
		achieving good throughput. We believe that once the pause times are
		optimised for, optimising for throughput is easier, but the vice versa is
		much harder.
\end{enumerate}

We develop a generational garbage collector with two generations where the old
generation (major heap) is shared between all of the mutators and is collected
with a non-moving, mostly-concurrent, mark-and-sweep collector modelled on
VCGC~\cite{Huelsbergen98}. VCGC avoids having explicit phase transitions
between marking and sweeping, which is the traditional source of bugs in
concurrent collector designs. For programs that do not use weak references or
ephemerons, the mutator and the GC threads need to synchronize only once
per cycle to agree that the current cycle is done. This minimizes the pause
times. The major heap allocator is based on the Streamflow~\cite{Schneider06}
design, which uses size-segmented thread-local pages. This has been shown to
have good multicore behaviour and fragmentation performance.

For the young generation (minor heap), survivors are copied to the shared major
heap. We present two alternative collectors for the minor heap with different
tradeoffs (\S\ref{sec:minor}). The first is a concurrent collector with
thread-private \emph{minor} heaps~\cite{Doligez93, Anderson10, Domani02,
Auhagen11, Marlow11, Sivaramakrishnan14}. The original quasi-real-time
collector design by~\cite{Doligez93} maintains the invariant that there are no
pointers from major to minor heaps. Thus, storing a pointer to a private object
into the shared major heap causes the private object and all objects reachable
from it to be promoted to the shared heap en masse. Unfortunately, this eagerly
promotes many objects that were never really shared: just because an object is
pointed to by a shared object does not mean another thread is actually going to
attempt to access it. A good example is a shared work-stealing queue, where
stealing is a rare operation. It would be unwise to promote all the work to the
major heap.

Our concurrent minor collector design (\S\ref{sec:concminor}) is similar but
lazier, along the lines of GHC's local heaps~\cite{Marlow11}, where objects are
promoted to the shared heap whenever another thread actually tries to access
them. This has a slower sharing operation, since it requires synchronisation of
two different threads, but it is performed less often. However, this design
requires that reads be safe points where garbage collection can occur. Consider
the case of two mutators both of which are trying to read an object in the
minor heap of the other mutator. In order to make progress, the mutators will
have to service promotion requests that they receive while waiting for their
request to be satisfied. Recall that stock OCaml does not use read barriers and
the C API also works under this assumption. By making the reads safe points, it
is likely that every user of the C API will need to update their code to
conform with the new semantics. This conflicts with our first requirement.

To this end, we develop a stop-the-world parallel minor collector
(\S\ref{sec:stwminor}) where all the mutators will need to synchronize to
perform the minor collection, at the cost of possible worse pause times. The
memory access invariants remains the same, and hence, no change is necessary
for the C API. One surprising result we will justify in our evaluation
(\S\ref{sec:eval}) is that the stop-the-world minor collector outperforms the
concurrent minor collector in almost all circumstances, even as we cranked up
the number of cores. This result gives us a clear choice in the design space to
pick a backwards-compatible {\em and} performant concurrent garbage collector.

\subsection{Contributions}

Our contributions are to present:

\begin{itemize}
	\item the design of a mostly-concurrent, non-moving, mark-and-sweep GC for
		the older generation that minimizes pause times for a parallel extension of
		OCaml.
	\item two collector designs for the young generation: {\em (i)} a
		concurrent collector that minimizes pause times at the cost of breaking the
		C API and; {\em (ii)} a stop-the-world parallel collector that retains the
		backwards compatibility of the existing C API.
	\item extensions of our baseline collectors to advanced language features
		that interact with the GC such as lazy values, finalisers, weak references
		and ephemerons. Our novel design minimizes the number of global
		synchronizations necessary for collecting a deeply nested hierarchy of
		ephemerons. This design has been verified in the SPIN model checker.
	\item support for \emph{fibers} that run in parallel, which are language
		level lightweight threads implemented as runtime managed stack segments.
		The implementation of fibers is similar to lightweight threads in Haskell
		GHC and Goroutines in the Go language. While the details of the language
		support for fibers is beyond the scope of the paper, we describe the subtle
		interaction of our concurrent GC algorithm with fibers.
	\item extensive evaluation of the collector designs in a full-fledged
		implementation of a parallel extension of OCaml. Our experiments illustrate
		that {\em (i)} serial programs retain their performance profile on the new
		collectors, and {\em (ii)} parallel programs achieve good multicore
		scalability while preserving low pause times with increasing number of
		cores.
\end{itemize}

The rest of the paper is organized as follows. In the next section, we give an
overview of memory management and garbage collection in OCaml, which applies to
both stock OCaml and our parallel extension. \S\ref{sec:major} gives a
detailed description of the major heap allocator and collector, and the changes
required to make it parallel, while \S\ref{sec:minor} does the same for the
minor heap. \S\ref{sec:awkward} describes the extension of the new GC with
lazy values, finalisers, weak references, ephemerons, and fibers.
\S\ref{sec:eval} presents extensive performance evaluation of the new GC
against the goals set in \S\ref{sec:req}. In the light of this performance
evaluation, \S\ref{sec:discussion} discusses the path ahead for retrofitting
parallelism onto OCaml. \S\ref{sec:related} and \S\ref{sec:conc} present the
related work and conclusions.

In the remainder of the paper, we call our parallel extension of OCaml
``Multicore OCaml'' to distinguish it from stock OCaml. Following standard GC
terminology, we refer to the user program as ``the mutator'', even though
actual mutation is relatively rare in OCaml programs.

\section{An Overview of Memory Management in OCaml}
\label{sec:stockOCaml}

OCaml uses a uniform memory representation in which each value has the same
size~\cite{leroy90zinc}, making it possible to compile just one copy of
polymorphic functions~\cite{appel90runtime}. Values are always one word long
(either 32 or 64 bits, depending on architecture), and consist of either an
integer or a pointer. The least significant bit (LSB) is used to distinguish
integers and pointers: since all allocations are word-aligned, the LSB of
pointers is guaranteed to be zero, whereas integers are encoded by
left-shifting the integer value by one, and setting the LSB to one. Hence,
integers in OCaml are 31- or 63-bits long. Every OCaml object has a word-sized
header, which encodes the length and type of the object, and has two bits
reserved for encoding the colours used by the major collector. (For more
details of OCaml's memory representation, see \citet[Ch. 20]{RWO}).

OCaml uses a generational GC with a major and minor heap~\cite{doligez89gc}.
The minor heap is small (256K words, by default), where new objects are
allocated by bumping the allocation pointer. The minor heap is collected when
it is full by a copying collector which copies live objects to the major heap.
The minor collection proceeds by first promoting all the roots (globals,
\emph{local roots} registered in C calls, registers, the \emph{remembered set} of
inter-generational pointers from the major to the minor heap, and the program
stack) that point to the minor heap to the major heap, and then transitively
promoting all the referenced objects. All references from live objects to
objects in the minor heap are updated to point at the new location, including
references from the major heap to the minor heap (recorded in the remembered
set), and the old minor heap is then reused in the next cycle. The copying
collector only needs to touch the live objects. Given that the survival rate in
the minor collection is low, a copying collector minimizes pause times.

The major heap contains objects which have survived a minor collection (as well
as objects above a certain size, which are allocated there directly). Instead
of a bump-pointer algorithm, allocation in the major heap uses a more
sophisticated allocator, which differs between stock and Multicore OCaml. The
major GC is incremental, except for an optional stop-the-world compaction
phase.

\section{Major heap}
\label{sec:major}

Next, we present the details of OCaml's major heap and garbage collector, and
the changes necessary to retrofit parallelism. OCaml's major collector is:
\begin{description}
\item[Mark-and-sweep] Collection is divided into two phases: \emph{marking}
	determines which allocations are still in use, and \emph{sweeping} collects
		those that are not and makes them available for reuse.
\item[Non-moving] Once allocated in the major heap, objects remain at the same
	address until collected.
\item[Incremental] Rather than stopping the program for the whole duration of
	GC (which would cause a pause bounded in length only by the size of memory),
		the OCaml major collector pauses the program in many small \emph{slices}.
\end{description}
Multicore OCaml's collector retains these properties, and is also
\emph{parallel}: the virtual machine contains a number of \emph{domains}, each
running as a separate system thread in the same address space. Domains can be
dynamically created and brought down.

A garbage collector uses a large amount of shared mutable state, as the
collector must track how each byte of memory is being used. The central
challenge in writing a multicore garbage collector is controlling access to
this state without introducing too much expensive synchronisation. To
understand this challenge, we first review how this state is used in the stock
OCaml collector.

\subsection{Tricolour collection in stock OCaml}

OCaml uses the standard tri-colour abstraction~\cite{dijkstra78gc}: allocations on
the major heap are given one of three colours \emph{black}, \emph{gray} and
\emph{white}. The colour of an object is stored in the two GC bits in the
object's header. The fourth possible value of these bits, named \emph{blue} by
OCaml, is used to indicate blocks that are not objects but free memory,
available for future allocations.

\begin{wrapfigure}{r}{0.65\textwidth}
	\includegraphics[scale=0.39]{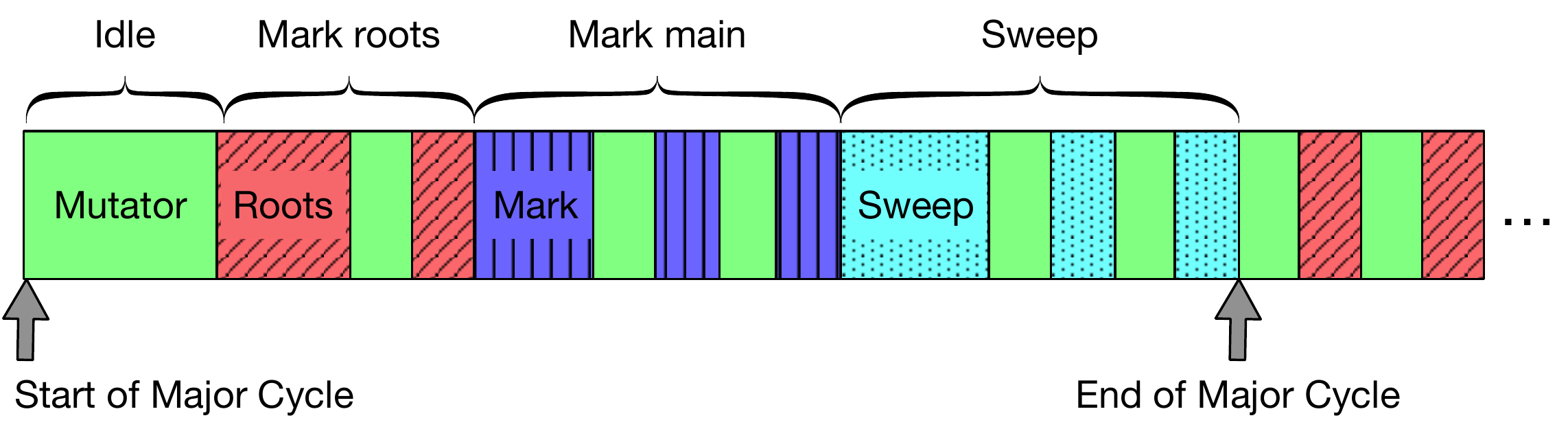}
	\caption{Major collection in stock OCaml}
	\label{fig:stock_gc}
\end{wrapfigure}
The collector operates in two phases, first \emph{marking} and then
\emph{sweeping}. These phases are incremental and so are interleaved with the
user program (Figure~\ref{fig:stock_gc}), which in particular may \emph{mutate}
the heap and \emph{allocate} new objects.

\paragraph*{Marking}
The goal of the marking phase is to colour all live objects black, leaving dead
(garbage) objects coloured white. The gray objects represent the frontier, that
is, those objects which have themselves been marked but which have not yet
been scanned for outgoing pointers.

The first part of marking is to mark the roots. Marking makes a white object
grey and pushes it into the mark stack. The root marking is itself split into
two phases -- a non-incremental phase where the registers and the program stack
are scanned, and an incremental phase for scanning the global roots. Given that
real world programs may have a large number of global roots, incrementally
marking them reduces pause times.

After the roots are marked, marking is continued by popping the mark stack,
marking the children of the object popped, and then marking the object black.
This process is repeated until the mark stack becomes empty, meaning that there
are no more gray objects.

\paragraph*{Sweeping}
When marking is completed, the objects that are still in use are black and the
garbage is white. The sweeping pass does a single traversal of the whole heap
(again, incrementally), turning black objects back to white and turning white
objects blue, to indicate that their storage is now free to be used for future
allocations.

\paragraph*{Mutation}
When mutating a field of an object, the program invokes the \emph{write
barrier}.  During the mark phase of the GC, the write barrier loads the object
pointed to by the \emph{old} value of the field, and if it is white, grays it
and adds it to the mark stack. This preserves the invariant that every object
reachable at the start of marking eventually gets marked (the
\emph{snapshot-at-the-beginning} property).

The write barrier also keeps track of the inter-generational pointers from
major to minor heap in a remembered set, which is used as a root for the minor
collection. (See \cref{sec:minor}).

\paragraph*{Allocation}
New allocations must be coloured so that they are not immediately
collected. During the marking phase, this means that they are coloured
black. During the sweeping phase, their allocation colour depends on whether
sweeping has yet reached their position: if so, they are coloured white, but if
not, they are coloured black so that the sweeper later turns them white.

\medskip

This is just one possibility from a large design space. OCaml grays the
\emph{old} value of a field during mutation (a \emph{deletion
  barrier}~\cite{Yuasa90}) and marks the roots \emph{before} marking anything
else (a \emph{black mutator}). This has the advantage that the total amount of
work to do during a GC cycle is bounded, regardless of what the mutator does
(guaranteeing termination of GC), but the disadvantage that anything allocated
on the major heap during the marking phase of one cycle can be collected at the
earliest a full cycle later. See \citet{vechev2005collectors} or
\citet{GCHandbook} for further discussion of these details. We will not discuss
them further here, as the trade-offs involved are largely independent of the
switch from a single-threaded to a multi-threaded GC.

What we do discuss here is the shared state involved in this design.  Mutable
state is shared between marking and sweeping (the object colours, whose meaning
changes between phases), between mutation and marking (the write barrier
supplies new gray objects to the collector) and between allocation and sweeping
(allocation must determine the phase of the collector and position of sweeping,
as well as coordinating with the sweeper to manage free memory).

In a single-threaded collector, organising access to such shared state is easy,
but doing so correctly and efficiently is the central challenge of a parallel
collector. Even the seemingly-simple state of whether the collector is marking
or sweeping is problematic: many designs have had errors in the case where the
GC transitions phase during a write barrier, relying on the precise order of
operations~\cite{gries77parallel}, which becomes even trickier under weak
memory~\cite{gammie2015tsogc}.

\subsection{Multicore OCaml's collector}

The approach we take in multicore OCaml is to avoid as much as possible of this
shared state. First, to reduce state shared between mutators and GC, we do not
use an explicit gray colour. Instead, each domain maintains its own stack of
gray objects, and when the write barrier needs to gray an object it marks it
and adds it to the local domain's gray stack. Since GC and OCaml code are
interleaved on a single domain, no synchronisation is necessary. (This scheme
does mean that it's possible for an object to be on the mark stack of two
separate domains. See below).

\begin{wrapfigure}{r}{0.6\textwidth}
	\includegraphics[scale=0.31]{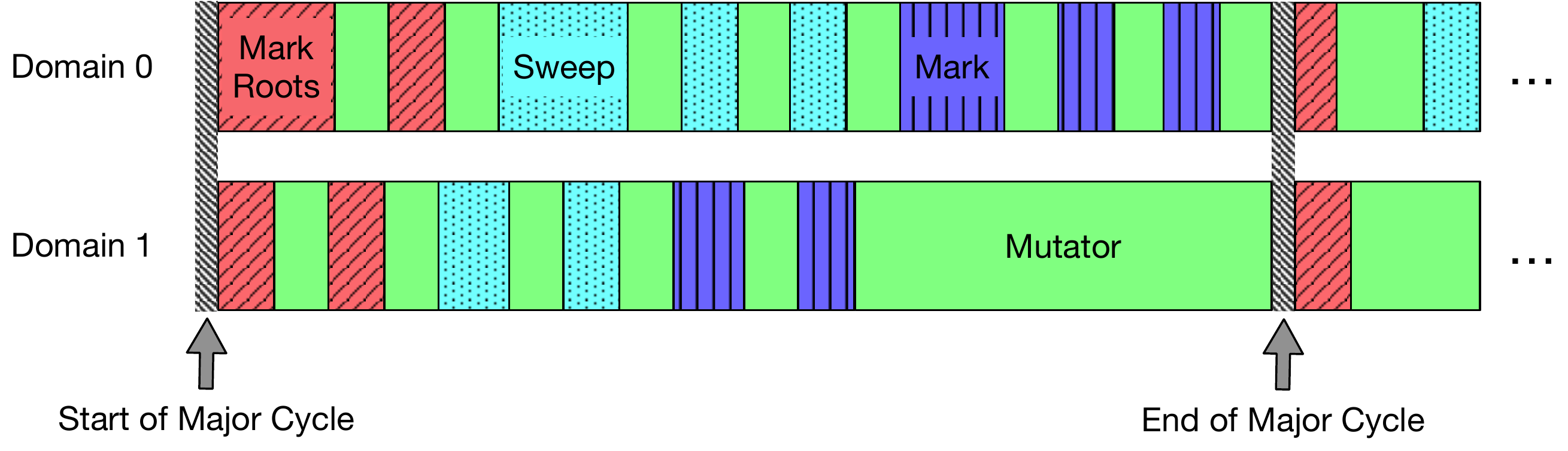}
	\caption{Major collection in Multicore OCaml. Our design permits the marking
	and sweeping phase to overlap and avoids the global synchronization between
	the phases.}
	\label{fig:multicore_major}
\end{wrapfigure}

Avoiding shared state between marking and sweeping is trickier, and to this end
we reuse a design from the Very Concurrent Garbage Collector
(VCGC)~\cite{Huelsbergen98}, which does not have distinct marking and sweeping
phases. Instead of colouring objects black and white (with the meaning of black
and white changing as we move between mark and sweep phases), Multicore OCaml
uses three states: |Marked|, |Unmarked| and |Garbage|. For regions of free
memory, there is a fourth state |Free|, corresponding to OCaml's blue.

Marking proceeds by changing |Unmarked| objects into |Marked| ones, while
sweeping proceeds by collecting |Garbage| objects into |Free| blocks. Note that
the sets of objects affected by marking and sweeping are disjoint, so no
synchronisation between them is necessary. Likewise, new allocations are always
|Marked|, ensuring that they will not be immediately collected. No
synchronisation with the collector is required to determine allocation colour.

As well as state shared between the various phases of the collector and the
mutator itself, in a parallel collector we must also think about state shared
between multiple domains doing the same phase of the collector in parallel. We
resolve this by making marking \emph{idempotent} and sweeping \emph{disjoint}.
Multiple domains may attempt to mark the same object at the same time, and we
make no attempt to avoid this. Instead, we allow the occasional object to be
marked twice, knowing that this gives the same result as marking once. Allowing
for this possibility is much cheaper than synchronising each access to an object
by the collector.

Sweeping is not idempotent. Instead, we ensure that the areas swept by different
domains are disjoint: each domain sweeps only the memory that it allocated,
keeping the interaction between sweeping and allocation local to a domain.

There is only one synchronisation point, at the end of the collection cycle.
When marking is completed (that is, all domains have empty mark stacks, leaving
only unreachable objects |Unmarked|) and sweeping is completed (that is, there
are no more |Garbage| objects, all having been made |Free|), all of the
domains simultaneously stop. With all domains stopped, we relabel the states:
the bit-patterns used to represent |Marked| now mean |Unmarked|, those for
|Unmarked| now mean |Garbage| and those for |Garbage| now mean |Marked|. This must
be done with all domains stopped, but is a small, constant amount of work, for
which the only difficulty is deciding when it should happen.

\subsubsection{Termination Detection}
\label{sec:termination}

Due to the use of deletion barrier and the fact that new objects are allocated
|Marked|, the amount of marking work in a given cycle is fixed
(snapshot-at-the-beginning property). Any object that was alive at the end of
the previous cycle will be marked in this cycle. Having fixed amount of mark
work is crucial to determine when the marking phase (and also the major cycle)
is done, especially when new domains may be created dynamically.

Domains may also be destroyed dynamically. Before exiting, a terminating domain
will sweep all of its local pages returning them to the global pool and keep
marking until its mark stack is empty.

\begin{figure}
	\begin{minipage}[t]{0.40\textwidth}
\begin{code}
def majorSlice (budget):
	budget = sweepSlice (budget)
	budget = markSlice (budget)
	if (budget && !dlMarkingDone):
		dlMarkingDone = 1
		atomic:
			gNumDomsToMark--
	if (gNumDomsToMark == 0):
		runOnAllDoms (cycleHeap)
\end{code}
	\end{minipage}
	\begin{minipage}[t]{0.58\textwidth}
\begin{code}
def cycleHeap ():
	barrier:
	/* run only by last domain to reach barrier */
		newc.Unmarked = gColours.Marked
		newc.Garbage = gColours.Unmarked
		newc.Marked = gColours.Garbage
		newc.Free = gColours.Free
		gColours = newc
		gNumDomsToMark = gNumDoms
	dlMarkingDone = 0
	markNonIncrementalRoots ()
\end{code}
	\end{minipage}
	\caption{Pseudocode for slice of major heap work. We use the prefix
	\cf{g} and \cf{dl} to distinguish global and domain-local variables.}
	\label{code:slice1}
\end{figure}

The pseudocode in Figure~\ref{code:slice1} illustrates the termination
detection and heap cycling algorithms. We use the prefix |g| and |dl| for
global and domain-local variables. |majorSlice| performs a slice of the major
GC work. The functions |sweepSlice| and |markSlice| take in a budget (in words)
and return the unspent budget. At the start of the cycle, we record the number
of active domains in the global variable |gNumDomsToMark|. If the call to
|markSlice| returns a non-zero value and the domain had some marking work to do
before the call, then it indicates that the mark stack on that domain has now
been emptied, which we record by decrementing |gNumDomsToMark|. In the absence
of ephemerons and weak references, this variable monotonically decreases to 0.

\subsubsection{Heap cycling}
\label{sec:heap_cycling}

The domain which completes marking last requests barrier synchronization to
complete the major cycle (|cycleHeap|). The last domain that enters the barrier
executes the code in the body of the |barrier| block. We swap the meaning of the
GC colours as described above, and record the number of domains at the barrier
(|gNumDoms|) as the number of domains to mark in the next cycle. Hence, the
stop-the-world phase is quite short in our GC design.

After the stop-the-world phase, the domains reset |dlMarkingDone| variable to
indicate that marking is not complete for the next cycle. The domains also mark
the non-incremental roots (local roots, current stack and register set) before
continuing with the mutator. Similar to stock OCaml, global roots are marked
incrementally as part of the regular marking work.

Any new domains that are created in a cycle will not have any marking work to
do in that cycle. This is because any of the new domain's roots will have to
(transitively) come from one of the domains that were running at the end of the
previous cycle, and will either have been marked by another domain or were
allocated already |Marked|. Hence, they are not included in |gNumDomsToMark|.

\if{0}
\subsection{Mark stack management}
Mark stack is only allowed to grow to 1/32nd of the size of the major heap.
Beyond this, the mark stack is pruned by dropping half of the entries, and the
heap is marked \emph{impure}. To mark an impure heap, the heap is first set to
\emph{pure}, and the blocks are scanned in increasing order of memory addresses
from the lowest addressed block. If the scan finds a grey object, then it
pushed to the mark stack and handled as usual. Once the scan is complete,
marking checks again that the heap is still pure. The process is repeated until
the mark stack becomes empty and the heap remains pure.
\fi

\subsection{Allocator}

Since all values allocated in OCaml are immediately initialised, a performance
goal for the allocator is to make the cost of allocation roughly proportional to
the cost of initialisation. In other words, small objects (lists, options,
etc.) must be allocated quickly, but larger allocations may be slower.

In stock OCaml, there is a choice of major heap allocators. The first-fit and
next-fit policies are classical single-free-list allocators, satisfying
allocation requests from the first sufficiently large free block or the next one
after the previous allocation, respectively. Recently, a best-fit allocator has
been incorporated into OCaml~\cite{BestFit} which uses segregated
free-lists~\cite{Standish80} for small object sizes (the common case), and a
splay tree~\cite{Sleator85} of blocks of different sizes for larger allocations.
Best-fit has been observed to beat the existing algorithms on both time spent
and space used through fragmentation.

However, all of these allocators are single-threaded. Due to the high
allocation rates in OCaml, using a lock to protect the allocator would have an
unacceptably high performance penalty. For this reason, Multicore uses a
different major heap allocator  based on the Streamflow~\cite{Schneider06}
design. Our allocator maintains a domain-local, size-segmented list of pages
for small allocations (less than 128 words). Each page of 4K words is carved
into equal-sized slots, with size classes chosen so that there is less than 10\%
wasted space.

Large allocations (at least 128 words) are forwarded to the system |malloc|,
maintained in a domain-local list of large blocks, and returned via system
|free| when swept. Before a domain
terminates, all of the pages it owns are moved to a global size-segmented list
of pages. Such pages are said to be \emph{orphaned} and must be serviced by the
other domains as they might contain live objects referenced by other domains.
The access to this global list of pages is protected by a mutex.

Streamflow uses BiBoP~\cite{Steele77} to track which slots in a page are free
without requiring object headers. However, since every OCaml object already has
a header, we use it instead to encode which slots are free.

In order to service an allocation request to the major heap, the domain searches
for a suitable slot in the corresponding size class in its local list of
pages. If no such slot is found, the domain sweeps the local unswept pages of
that size class to find a free slot. Failing that, the domain \emph{adopts} one
of the global pages of the appropriate size class with an available
slot. Failing that, the domain adopts one of the full global pages of the
appropriate size class, sweeps it, and looks for a free slot. If that case fails
too, the domain will request the operating system to allocate an additional
page. Thus, most allocations can be serviced without the need for
synchronization and atomic operations.

\subsection{Safe points}

In order to support the stop-the-world pauses required by major collector cycle
changes and the parallel minor collector, multicore OCaml uses \emph{safe
points} in addition to the existing allocation points. This is implemented
using the algorithm from \cite{Feeley93} and bounds the number of instructions
until a safe point is encountered. These safe points are necessary to avoid
deadlocks in certain situations e.g one domain in a non-allocating loop
spin-waiting on an empty queue while all other domains need a minor collection
in order to allocate new items to put on the queue. It should be noted that
safe points are required by stock OCaml for correct language semantics
independent of multicore.

\subsection{Opportunistic work}

The incremental, concurrent nature of the major collector allows work to be
carried out \emph{opportunistically} when domains might otherwise be idle or in
a state where it is unsafe for mutators to be running. Multicore OCaml makes use
of this property of the major collector to schedule opportunistic marking and
sweeping work in the following cases:
\begin{description}
\item[Yielding] Many non-blocking algorithms involve some form of looping when
attempting to make progress in the face of contention and involve some relaxing
e.g |PAUSE| on x86. We use this as an opportunity to do a small amount of
opportunistic marking or sweeping.
\item[Stop-the-world entry waits] In the parallel minor collector domains entering
a stop-the-world pause will carry out opportunistic work until all domains are
ready.
\end{description}

There are additional areas where opportunistic work could be implemented in
Multicore OCaml in the future, such as via read faults or stop-the-world leave
waits, but these are left as future work.

\section{Minor Heap}
\label{sec:minor}

The minor heap is much smaller than the major heap, but must sustain a much
higher rate of allocation and collection. Unlike the major collector, the minor
collector is:
\begin{description}
\item[Copying] At collection time, live objects in the minor heap are moved to
  the major heap. The minor heap is emptied by this process, and its space is
  reused in the next cycle.
\item[Non-incremental] The mutator is paused for the entire duration of the
  minor collection.
\end{description}
As in the previous section, we first review OCaml's single-threaded algorithm.

\subsection{Copying minor collection in stock OCaml}

Like the major collector, the minor collector traces objects from the roots to
find the live objects. Unlike the major collector, the set of roots also
includes the \emph{remembered set}, which consists of the references from the
major heap to the minor heap. This way, all objects in the minor heap that are
pointed to by something in the major heap at the time of minor collection get
promoted, without having to traverse the major heap to find them.

The number of objects to be copied by the minor collector is often small. This
is for two reasons: first, the minor heap is relatively small, and second, only
that part of the minor heap that is live at collection time must be traversed.
The limited work needed by the copying collector enables low pause times to be
achieved.

Since objects in the minor heap will be discarded or moved soon, there is no
need to use a sophisticated allocator to place them. Allocation is therefore
fast with a simple bump-pointer allocator.

\subsection{Parallelising the minor collector}

Compared to the major collector, the minor collector is more difficult
to parallelise as it moves objects, and an object must not be moved by
one thread while another is using it.

It is possible to parallelise a copying collector by carefully
coordinating the collector and the mutator, sharing state about which
objects have been moved. Such an approach comes at a heavy price,
however, in either performance or implementation cost, as it requires
fine-grained synchronisation between the collector and mutator not
just on mutations but on all memory accesses.

So, we avoid designs which involve copying objects in parallel with
their use by the mutator. This leaves two possibilities: to
separate the minor collector from the mutator in \emph{time} or in
\emph{space}.
\begin{description}
\item[Separation in time] When the minor heap fills, stop all domains
  simultaneously, and have all domains collect the minor heap in
  parallel before resuming the mutator.
\item[Separation in space] Give each domain a private minor heap,
  preventing any access by one domain to another's heap, and allow
  them to collect their minor heaps independently.
\end{description}
We have implemented both of these approaches, as described
below and evaluated in \cref{sec:eval}.

\subsection{Concurrent minor collector with private minor heaps}
\label{sec:concminor}

Our concurrent minor collector uses domain-local, private, minor heap
arenas, each tracking their own remembered set.
We maintain the invariant that there are no pointers between the minor heaps of
different domains, which permits each of the minor heaps to be independently
collected. We do allow pointers from the shared major heap to the minor heap to
avoid paying the cost for early promotion, similarly to multicore
GHC~\cite{Marlow11}.

\subsubsection{Read faults and interrupts}

Whenever a read follows a pointer from major to a remote minor heap, it
triggers a \emph{read fault}. The faulting domain issues an interrupt to the
remote domain to request promotion of the desired object and waits for the
response. Each domain maintains a multiple-producer single-consumer queue for
receiving inter-domain interrupts. The sender pushes the request into the
target domain's queue and modifies the minor heap allocation limit pointer on
the target domain such that the next allocation on the target domain would
fail.  This enables timely handling of inter-domain interrupts. We use the same
interrupt mechanism for signalling to other domains when a stop-the-world phase
is necessary at the end of the major cycle (\S\ref{sec:heap_cycling}).

When the target domain receives a promotion request, it promotes the transitive
closure of the requested object to the major heap (possibly triggering a minor
GC on this domain), and returns the location of the new object in the major
heap to the source domain. Since two domains may concurrently request promoting
objects from each other's domains, while the faulting domain waits for a
response from the remote domain, it polls its interrupt queue and handles
requests.

\subsubsection{Read Barrier}
\label{sec:readbarrier}

\begin{wrapfigure}{r}{0.45\textwidth}
	\centering
	\includegraphics[scale=0.4]{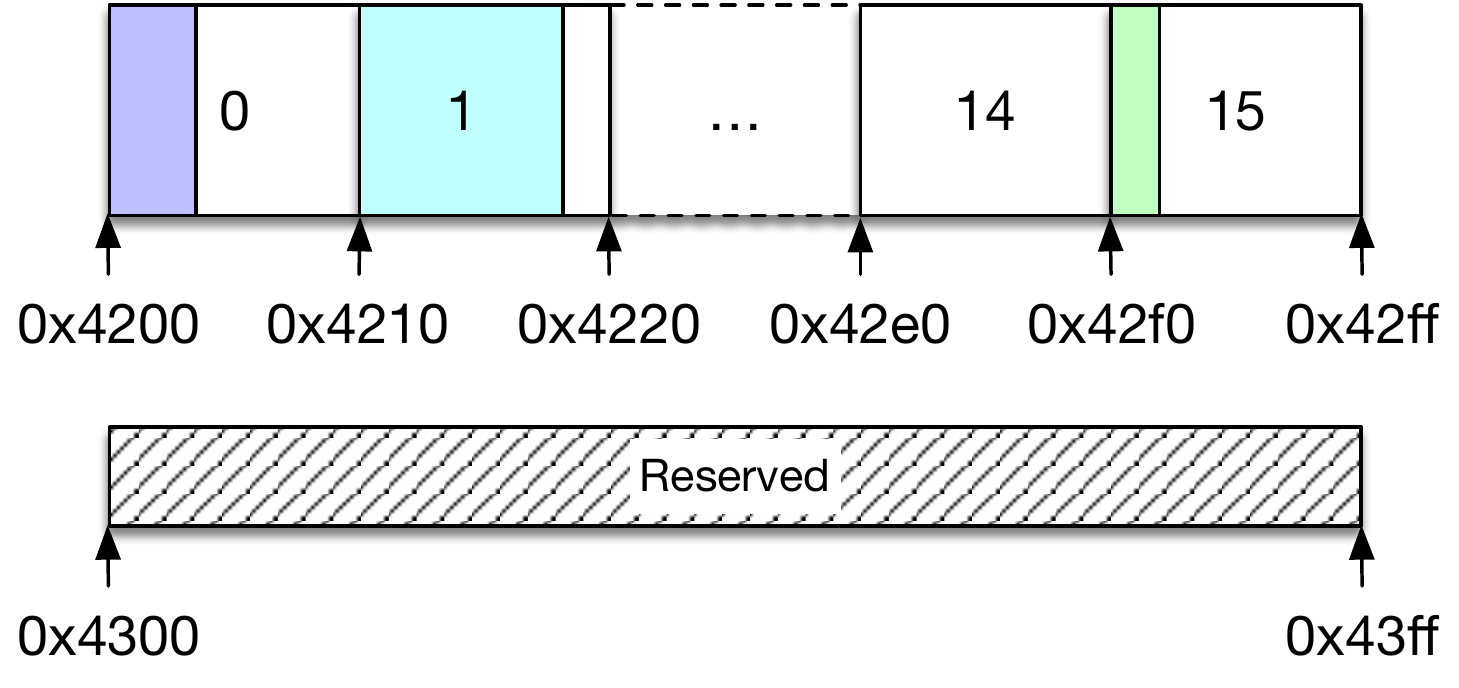}
	\caption{An example virtual memory mapping of the minor heaps for the
	concurrent minor collector on a 16-bit address space.}
	\label{fig:multicore_minor_vmm}
\end{wrapfigure}
The main challenge in the concurrent GC design is optimising the read barrier,
which is a code size and execution time burden not present in stock OCaml.

There is an efficient way to implement the read barrier
through careful virtual memory mapping for minor heap arenas and bit
twiddling. When reading a value from a mutable location, the read barrier must
classify the value as:
{\em (a)} an integer; or {\em (b)} a value in the major heap; or {\em (c)}
a value in its own minor heap; or {\em (d)} a value in a remote minor heap.
We must distinguish the last case from the others, as it requires a read fault.

\begin{wrapfigure}{l}{0.45\textwidth}
	\begin{code}
	/* ax = value of interest */
	/* bx = allocation pointer */
	xor %bx, %ax
	sub 0x0010, %ax
	test 0xff01, %ax
	/* ZF set => ax in remote minor */
	\end{code}
	\caption{Read barrier test}
	\label{code:read_barrier}
\end{wrapfigure}
Figure~\ref{fig:multicore_minor_vmm} shows an example
layout of the minor heap arenas, using only 16-bit addresses for clarity.
The minor heaps are all allocated in a
contiguous power-of-2 aligned virtual memory area, where each minor heap is
also a power-of-2 aligned and sized. Not all of the virtual memory area needs
to be allocated, but only needs to be reserved to prevent it being used for
the major heap. In the figure, the shaded
regions indicate committed memory currently used for minor heap arenas.

In this example, the minor heap arenas are all carved out of the virtual memory
space between |0x4200| and |0x42ff|. We also reserve the virtual
memory space |0x4300| to |0x43ff|, and allocate neither minor or
major heap pages in this space. We have chosen a layout with support for 16
domains, each with a maximum minor heap size of 16 bytes. Addresses in this
16-bit address space can be written as 4 quads |0xPQRS|. In OCaml,
integer values are represented by tagging the least significant bit to be 1.
Hence, in this example, integers have low bit of |S| to be 1. Minor heap
values have |PQ| to be 42, and |R| determines the domain.

We can implement the read barrier check by comparing the given address with an
address from the minor heap. Luckily, we have such an address handily available
in a register-- the minor heap allocation pointer. Let us assume the x86
architecture, with the allocation pointer in the |bx| register, and the value
that we want to test in |ax|. The read barrier test is given in
Figure~\ref{code:read_barrier}. The |test| instruction performs bitwise |and|
on the operands, and sets the zero-flag (|ZF|) if the result is 0. At the end
of this instruction sequence, if the zero-flag is set, then |ax| is in a remote
minor heap. To see how this works, let us consider each of the four cases
(Figure~\ref{code:read_barrier_proof}).

\begin{figure}
  \begin{subfigure}[b]{0.38\textwidth}
    \begin{code}
      /* low_bit(ax) = 1 */
      /* low_bit(bx) = 0 */
			xor %bx, %ax
      /* low_bit(ax) = 1 */
			sub 0x0010, %ax
      /* low_bit(ax) = 1 */
			test 0xff01, %ax
      /* ZF not set */
		\end{code}
    \caption{Case: Integer}
    \label{code:rb_int}
  \end{subfigure}
  \begin{subfigure}[b]{0.58\textwidth}
    \begin{code}
      /* PQ(ax) = 0x42 */
      /* PQ(bx) != 0x42, PQ(bx) != 0x43 */
			xor %bx, %ax
			/* PQ(ax) != 0, PQ(ax) != 1 */
			sub 0x0010, %ax
      /* PQ(ax) != 0 */
			test 0xff01, %ax
      /* ZF not set */
		\end{code}
    \caption{Case: major heap}
    \label{code:rb_major}
	\end{subfigure}
  \begin{subfigure}[b]{0.38\textwidth}
    \begin{code}
      /* PQR(bx) = PQR(ax) */
			xor %bx, %ax
      /* PQR(ax) = 0 */
			sub 0x0010, %ax
      /* PQ(ax) = 0xff */
			test 0xff01, %ax
      /* ZF not set */
		\end{code}
    \caption{Case: Own minor heap}
    \label{code:rb_my_minor}
  \end{subfigure}
  \begin{subfigure}[b]{0.58\textwidth}
    \begin{code}
      /* PQ(bx) = PQ(ax) */
      /* lsb(bx) = lsb(ax) = 0 */
      /* R(bx) != R(ax) */
			xor %bx, %ax
      /* R(ax) != 0 */
      /* PQ(ax) = lsb(ax) = 0 */
			sub 0x0010, %ax
      /* PQ(ax) = lsb(ax) = 0 */
			test 0xff01, %ax
      /* ZF set */
		\end{code}
		\caption{Case: Remote minor heap. \cf{lsb} returns least significant bit.}
    \label{code:rb_remote_minor}
	\end{subfigure}
\caption{Cases of the read barrier.}
\label{code:read_barrier_proof}
\end{figure}
Consider the case when |ax| contains an integer
(Figure~\ref{code:rb_int}). Since the allocation pointer is always word aligned
in OCaml, the least significant bit of the allocation pointer |bx| will
be 0. Hence, zero flag will not be set at the end of the instruction sequence.

Consider the case when |ax| contains a major heap pointer
(Figure~\ref{code:rb_major}). Due to our virtual memory layout for minor heaps
which reserves additional unused virtual memory area adjacent to the used minor
heap area, the most significant eight bits in |bx| (written |PQ(bx)|), will
neither be |0x42| nor |0x43|. Hence, |PQ(ax)| after |xor| will be non-zero.
Hence, zero flag will not be set at the end of the instruction sequence.

Consider the case when |ax| contains a pointer to the domain's own minor
heap (Figure~\ref{code:rb_my_minor}). In this case, |PQR(ax) = PQR(bx)|.
After |xor|, |PQR(ax)| is 0. The subsequent subtraction
underflows, and hence, the |test| does not set the zero flag.

Consider the case when |ax| contains a pointer to a remote minor heap
(Figure~\ref{code:rb_remote_minor}). In this case, |PQ(ax) = PQ(bx)|,
the least significant bits (|lsb|) of |ax| and |bx| are 0,
and |R(ax) != R(bx)|. Hence, after |xor|, |R(ax) != 0|,
|PQ(ax) = lsb(ax) = 0|. Subtracting 1 from the non-zero |R(ax)|
will not underflow. Hence, |PQ(ax)| and |lsb(ax)| are still 0
after subtraction, and the zero flag is set. At this point, the domain raises
the interrupt to the target domain to promote the object to the major heap.

Our implementation allows the maximum size of the minor heap and the maximum
number of minor heap arenas (and hence, the number of domains) to be configured
at the compiler build time. On 64-bit architectures, by default, the compiler
is configured to host a maximum of 128 domains, each with a maximum minor heap
size of 16 MB. This necessitates reserving (not allocating) 4GB of virtual
address space, which is a small portion of that available on a 64-bit machine.

\subsubsection{Promotion}
\label{sec:promotion}

When a domain receives a promotion request, how should it go about
doing it? Because of mutable fields (and finalisers), it is not valid
to duplicate the requested object closure in the major heap, and we must
ensure all references are updated to the promoted version.

A correct but inefficient implementation would be to perform a minor garbage
collection, which will ensure that the requested object and everything else in
the minor heap is promoted to the major heap. This strategy suffers from early
promotion, and both the requester and the target domain will suffer the pause
time for a full minor GC on the target domain.

Observe that we can fix the early promotion problem by promoting only the
transitive closure of the requested object, and then scanning the minor GC
roots and the minor heap for pointers to the promoted objects and forwarding
them to the major heap copy. However, this solution is also inefficient, as it
would touch the entire minor heap (recall that the copying collector need only
touch live objects during a minor GC, a small portion of the heap).

Our solution is a hybrid of the above techniques. In our experiments, we
observe that most promoted objects were recently allocated (as was observed in
the case of a parallel extension of the MLton Standard ML
compiler~\cite{KC12}). We enhance the write barrier to record in a domain-local
\emph{minor remembered set} the writes from old to younger minor heap objects.
How do we identify whether an intra-minor-heap pointer is from an old to a
younger object? In OCaml, the allocation pointer starts at the end of the minor
heap area, and objects are allocated in the minor heap by subtracting the
allocation pointer by the required size. During a write |r := v|, if both |r|
and |v| are objects in the minor heap, and |r| is at a higher address than |v|,
then |r| is older than |v|. We add |r| to the minor remembered set.

When an object closure being promoted is recently allocated (in the last 5\% of
the minor heap), we promote the object closure. The pointers to those promoted
objects may only appear in either the minor GC roots, the minor remembered set,
or one of the objects allocated after the oldest promoted object, which is by
definition in the last 5\% of the minor heap. We scan those data structures
after object promotion. In our experiments, 95\% of the promoted objects were
the youngest 5\% of objects. If the object closure being promoted is not one of
the youngest 5\%, we perform a full minor collection. The minor remembered set
is cleared at the end of a minor collection.

\subsubsection{Rewriting remembered set}

In Multicore OCaml, the domain-local remembered set records the pointers from
the shared major heap to the minor heap, which must be updated during minor
collection. Consider that a major heap object |r| points to the
minor heap object |m|. After promoting |m|, |r| should be
rewritten to point to the new location of |m| in the major heap, say
|m'|. However, care must be taken to ensure that this write by the minor
GC does not overwrite a concurrent write to |r| by a different domain.
To this end, we perform the write to |r| by the minor GC with an
unconditional atomic compare-and-swap |CAS(r,m,m')|, which will fail if
|r| does not contain |m|. If it fails, then |r| has been written
to by a different domain, and hence, we ignore the failure.

\subsection{Stop-the-world parallel minor collector}
\label{sec:stwminor}

In this section, we describe the stop-the-world parallel minor collector which
is an alternative to the concurrent minor collector. Our stop-the-world
parallel minor collector retains domain-local minor heaps but relaxes the
invariant disallowing references between minor heaps. Inter-heap references
requires a stop-the-world synchronisation involving all domains to carry out
a minor heap collection. With stop-the-world minor collection, we do
not need read barriers and the read barriers need not be safe points. Hence, we
can retain the stock OCaml's C API.

\subsubsection{Interrupts and synchronization}

When a domain finds that it needs to collect its minor heap, it utilizes the
interrupt mechanism to signal that all domains should do a stop-the-world
collection. Care is taken to ensure that stop-the-world minor and major
collection requests are serialized so that only one can be executing at a given
time.

Given the increase in the frequency of interrupts, safe points become
necessary for the stop-the-world minor collector to avoid long synchronization
times before a minor collection can start. When all domains have entered the
interrupt a barrier is used to agree that a minor collection can start; at this
point we know that all the blocks in the minor heap can be safely moved.

\subsubsection{Parallel promotion}

Within the stop-the-world section, all domains promote reachable objects in
parallel. Care is needed to ensure that if two domains attempt to promote an
object, they are serialized using |CAS| operations on the object header. This
is a little more involved than a single |CAS| due to the need to update the
object's first field to forward to the new location in the major heap. Other
domains must never see a minor heap object with a zeroed header (indicating
promoted) before the first field is updated. In order to avoid this, the header
is |CAS|ed to an \emph{intermediate state} which indicates that a domain is
currently promoting the object. If a domain sees this \emph{intermediate state}
when attempting to promote an object they spin until the header is set to |0|,
at which point the first field now contains the forwarding pointer to the
promoted object in the major heap.

Additionally, we incorporate fast paths for the case when there is only one
domain running, which elides the |CAS| and the intermediate states. The minor
collection algorithm in the case when only one of the domains is running
involves the same steps as stock OCaml minor collection.

\subsubsection{Parallel work sharing}

We implement a static work sharing policy for speeding up the stop-the-world
remembered set promotion. This is necessary because each domain maintains its
own remembered set and there can be workloads where this leads to imbalance e.g
only one domain continually updates a set of globals with newly allocated data.
The static work sharing policy equally splits the remembered set entries across
all domains participating in the minor collection and aims to ensure each one
has an equal number of roots to promote.

\if{0}
\begin{figure}
  \begin{minipage}[c]{0.45\textwidth}
    \centering
    \begin{code}
      /* M = size of remembered set */
      /* N = number of domains */
      /* d = domain index */
      /* base = remembered set array */
      per_domain = M / N;
      start = base + (d * per_domain);
      if (d < N-1)
        end = base + ((d+1) * per_domain);
      else
        end = base + M;
    \end{code}
    \caption{Remembered set static sharing}
    \label{code:remembered_static_sharing}
  \end{minipage}
  \hfill
  \begin{minipage}[c]{0.45\textwidth}
    \centering
    \includegraphics[scale=0.4]{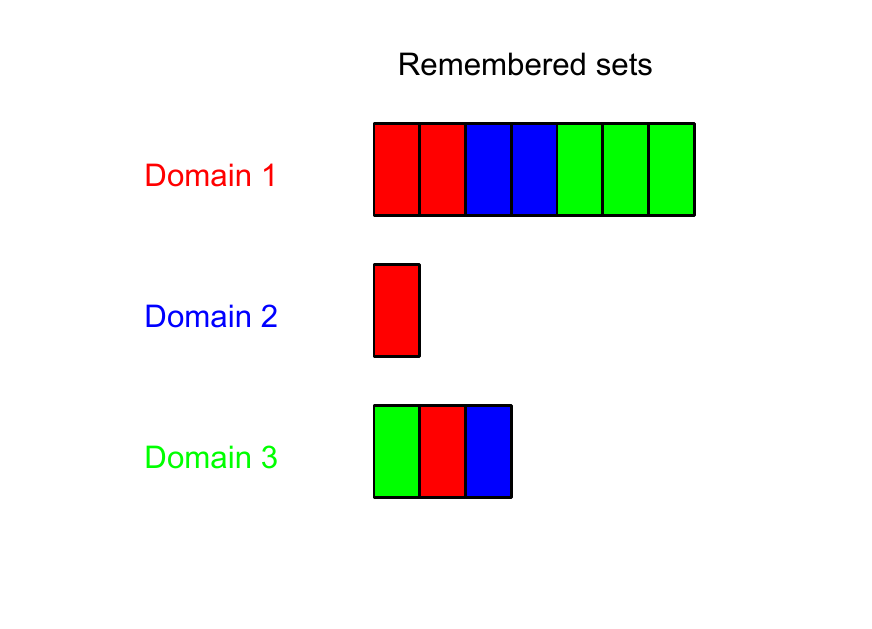}
    \caption{Example of remembered set static sharing}
    \label{fig:remembered_static_sharing_example}
  \end{minipage}
\end{figure}
\fi

Static work sharing only balances the \emph{roots} in the remembered set. It is
entirely possible that one domain may end up needing to promote a large object
graph reachable from one of its roots while other domains remain idle. This
could be addressed with dynamic work sharing, though it is unclear whether the
benefits would outweigh additional synchronisation between domains.

\section{Completing OCaml Language Coverage}
\label{sec:awkward}

OCaml has a substantial existing base of production code that has been
developed with single-core performance in mind.  Some of the popular areas the
language is used involve symbolic computation (theorem provers, proof
assistants, static analysis, etc.) and these will take some years to migrate to
parallel execution. Therefore, our decisions about how to retrofit parallelism
have been heavily influenced by the need to preserve fast single-core
performance, as it is impractical to maintain multiple runtimes.

We will now describe the extensions of our GC model needed to handle
the complete set of features in OCaml that interface with the GC, paying
careful attention to single-core performance to minimise impact on
existing code.

\subsection{Weak References and Ephemerons}

Weak references introduce the notion of \emph{weak reachability} in the heap.
An object |x| is said to be weakly reachable if it is referred to by a
weak reference. If |x| is not also strongly reachable (through normal
pointers), the garbage collector may collect |x|. The weak reference is
said to be full if it points to a value, and empty if the value was erased by
the GC.

An ephemeron~\cite{Hayes97,Bobot14} is an extension of a weak reference. It
consists of a key and a value. The ephemeron weakly points to the key. The
value is considered reachable if the ephemeron is reachable \emph{and} the key
is strongly reachable. Observe that ephemerons provide the mechanism to express
conjunction of reachability relations, whereas the usual GC mechanism can only
express disjunctions. If the key is not strongly reachable, the GC erases the
reference to the key and the value. We call this process \emph{ephemeron
sweeping}. OCaml implements weak references as ephemerons with only the key but
no value. Hence, we will only discuss ephemerons for the remainder of this
section. Ephemerons in OCaml may have more than one key, and the behaviour is
that, for the value to be considered reachable, the ephemeron and all of the
keys are strongly reachable.

We would like the ephemeron marking and sweeping to also not require
synchronizing all of the mutators and GC threads. To this end, each domain
maintains a list of ephemerons it has created. This list is incrementally
marked and swept by the owning domain \emph{concurrently} with the execution on
other domains. During domain termination, the ephemeron list is appended to a
global ephemeron list which is adopted by one of the running domains.

\subsubsection{Challenges}

Ephemerons bring in several challenges for a concurrent GC. Since the
reachability of ephemeron values can only be decided based on whether the keys
are strongly reachable, ephemeron marking must be performed after all the
domains have finished marking. During ephemeron marking, the domain local list
of ephemerons is walked, and for each ephemeron, if the ephemeron and its key
are strongly reachable, then the value is marked. Thus marking the ephemeron
may introduce additional marking work. This complicates our termination algorithm
(Section~\ref{sec:termination}), as the number of domains with marking work no
longer monotonically decreases. Moreover, the ephemeron value
when marked may make other ephemerons reachable, which may be on the ephemeron
list of another domain. Thus, if ephemeron
marking introduces new mark work, we will need to go through another round of
ephemeron marking.

OCaml ephemerons provide a |get_key| function which returns a strong reference
to the key if the key is full (not erased by the GC). This strong reference is
obtained by marking the key. If there were no other strong references to the
key, then the |get_key| operation introduces additional mark work in the
current cycle. Hence, just running the mutator may bring in additional mark
work and cause a domain to need to mark even though it had previously finished
the round's mark work. Only after all of the domains have marked their
ephemerons and have not produced any new mark work can we proceed to the
ephemeron sweep phase.

\begin{figure}
\begin{code}
/* Assume: pushing an object into an empty mark stack will increment
	[gNumDomsToMark] and set [dlMarkingDone = false]. */
\end{code}
	\begin{minipage}[t]{0.45\textwidth}
\begin{code}[numbers=left]
def majorSlice (budget):
	budget = sweepSlice (budget)
	budget = markSlice (budget)
	if (budget && !dlMarkingDone):
		dlMarkingDone = 1
		atomic:
			gNumDomsToMark--
			gEpheRound++
			gNumDomsMarkedEphe = 0
	/* Ephemeron Mark */
	cached = gEpheRound
	if (cached > dlEpheRound):
		budget=markEphe(budget,cached)
		if (budget && dlMarkingDone)
			dlEpheRound = cached
			atomic:
				if (cached == gEpheRound):
					gNumDomsMarkedEphe++
	/* Ephemeron Sweep */
	if (gPhase == SWEEP_EPHE):
		budget = sweepEphe(budget)
		if (budget && !dlSweepEpheDone):
		  dlSweepEpheDone = 1
			atomic: gNumDomsSweptEphe++
	/* Change Phase */
	changePhase ()
\end{code}
\end{minipage}
	\begin{minipage}[t]{0.50\textwidth}
\begin{code}[numbers=left,firstnumber=27]
def changePhase ():
	if (gPhase == MARK && gNumDomsToMark == 0
			&& gNumDomsMarkedEphe == gNumDoms):
		barrier:
			if (gPhase==MARK && gNumDomsToMark==0
       && gNumDomsMarkedEphe==gNumDoms):
				gPhase = SWEEP_EPHE
	if (gPhase == SWEEP_EPHE &&
			gNumDomsSweptEphe == gNumDoms):
		runOnAllDoms (cycleHeap)

def cycleHeap():
	barrier:
		newc.Unmarked = gColours.Marked
		newc.Garbage = gColours.Unmarked
		newc.Marked = gColours.Garbage
		newc.Free = gColours.Free
		gColours = newc
		gNumDomsToMark = gNumDoms
		gEpheRound = gNumDomsMarkedEphe = 0
		gNumDomsSweptEphe = 0
		gPhase = MARK
	dlMarkingDone = dlEpheRound = 0
	dlSweepEpheDone = 0
	markNonIncrementalRoots ()
\end{code}
\end{minipage}
\caption{Pseudocode for a slice of major heap work with ephemerons.}
\label{code:slice2}
\vspace{-0.3cm}
\end{figure}

\subsubsection{Ephemeron marking}

We extend the algorithm presented in Figure~\ref{code:slice1} to handle
ephemeron marking (Figure~\ref{code:slice2}). Due to the strict ordering
requirements between ephemeron marking and sweeping, we introduce two phases in
the GC -- |MARK| and |SWEEP_EPHE|. If a program does not use ephemerons,
then the |SWEEP_EPHE| phase can be skipped. The key idea is this: we say
that an ephemeron marking round is complete when a domain empties its mark
stack. If all of the domains are able to mark the ephemerons in the same round
and there is no new marking work generated, then the ephemeron marking phase is
completed. Importantly, our solution does not require synchronization between
each round of ephemeron marking.

For the sake of exposition, we assume that the number of domains
(|gNumDoms|) remains constant. The global |gNumDomsMarkedEphe| tracks the
number of domains that have marked their ephemerons in the current round. The
global |gNumDomsSweptEphe| tracks the number of domains that have swept
their ephemerons. The domain-local variable |dlSweepEpheDone| records
whether the current domain has swept its ephemerons. We maintain the global
ephemeron round in |gEpheRound|, and the last round for which a domain has
marked its ephemerons in |dlEpheRound|. All of these variables are
initialised to 0 at the start of the major cycle (see |cycleHeap|).

Whenever a domain empties its mark stack, it increments |gEpheRound| and
sets |gNumDomsMarkedEphe| to 0 to indicate that no domains have marked
ephemerons in the new round (lines 8--9). Before marking its ephemerons, a
domain caches a copy of |gEpheRound| (line 11). If it completes marking and
no new mark work has been generated, then it records that it has completed
marking for the cached round (line 15). If another domain had not concurrently
emptied its mark stack, then the cached round will be the same as the global
round, and the domain increments the global count of domains that have
completed marking in this current ephemeron round (lines 16--18).

\subsubsection{Termination}

\sloppy
How do we know that marking is done? In each major cycle, |gNumDomsMarkedEphe|
either increases from 0 to |gNumDoms| or is reset to 0. Because of
|Ephemeron.get_key|, |gNumDomsMarkedEphe == gNumDoms| and |gNumDomsToMark == 0|
can both be simultaneously true, however, running the mutator may create new
mark work. This increments |gNumDomsToMark|, which will in turn necessitate an
additional round of ephemeron marking. Hence, we check the termination
condition again in a global barrier to verify that new mark work has not been
created before switching to the |SWEEP_EPHE| phase (lines 28 -- 33). We have
verified the implementation of the major slice work and termination in SPIN
model checker~\cite{Holzmann97}.

\subsubsection{Ephemeron sweeping}

Sweeping ephemerons is comparatively simple. Each domain sweeps its ephemerons,
clearing keys and values if necessary and increments |gNumDomsSweptEphe|. This
variable monotonically increases to |gNumDoms|, at which point we cycle the major heap.

\subsection{Finalisers}

Each domain maintains a domain-local list of finalisers that it has installed,
and handles finalization locally. OCaml has two variants of finalisation
function:

\begin{code}[language=caml,keywords={val}]
val finalise : ('a -> unit) -> 'a -> unit
val finalise_last : (unit -> unit) -> 'a -> unit
\end{code}

|finalise| applies the higher-order function on the object being finalised, and
the function may revive the object by storing it in a global variable. Hence,
as part of finalisation, the object is marked. Hence, these finalisers are
marked to run when |gNumDomsToMark| first goes to 0, but before any of the
domains start marking ephemerons. Due to |Ephemeron.get_key|, we need a barrier
to move from the main marking phase to marking finalisers.

The |finalise_last| finalisers do not get hold of the object being finalised.
OCaml finalises them after all the marking has been completed for this major
cycle i.e., after we enter |SWEEP_EPHE| phase. Similar to the other cases, we
maintain a global variable each to keep track of the number of domains that
have run their |finalise| and |finalise_last| finalisers. These variables are
initialised to 0 and monotonically increase to |gNumDoms|. These variables are
consulted before moving to the next corresponding GC phase. This mirrors the
stock OCaml behaviour.
\ifdefined\techreport
The pseudocode for major slice with ephemerons and finalisers is given in
Appendix A.
\else
The pseudocode for major slice with ephemerons and finalisers can be found
in the Appendix A of the technical report~\cite{techrep}.
\fi

\subsection{Lazy values}

OCaml has support for deferred computations through the use of lazy values,
which interact with the GC. In stock
OCaml, a lazy value \ml{'a Lazy.t} has one of the following representations:

\begin{itemize}
	\item A block of size 1 with |Lazy| tag with a closure of type \ml{unit ->
		'a} in the first field.
	\item A block of size 1 with |Forward| tag with the computed value of type
		\ml{'a} in the first field.
	\item The computed value of type \ml{'a}\footnote{We elide a subtlety with
		floats for the sake of exposition}.
\end{itemize}

When a lazy value with |Lazy| tag and closure |f| is forced, the field is first
updated to \ml{fun () -> raise Undefined}, so that recursive forcing of a lazy
value raises an exception. Then, the deferred computation |f| is evaluated. If
|f| raises an exception |e|, then the computation is replaced with
\ml{fun () -> raise e}. Otherwise, the mutator modifies the header to |Forward|
and updates the first field to the result. Whenever the GC finds a reference to
a |Forward| object, it may short-circuit the reference to directly point to the
object.

This scheme is incompatible with a concurrent GC since the mutator modifying
the word sized header may race with a GC thread which is marking the object,
leading to one of the writes being lost. Moreover, the writes to update the
header to |Forward| and the field to the result of the deferred computation are
non-atomic. Hence, one of the two domains concurrently forcing a lazy may
witness one of the writes but not the other, leading to violation of type
safety (segmentation fault). Moreover, given that the deferred computation may
perform arbitrary side effects, it would be undesirable to duplicate the
effects when two domains concurrently force a lazy value.

For Multicore OCaml, we have redesigned lazy values to make them safe for
concurrent mutators and GC threads. We introduce a |Forcing| tag to indicate a
lazy value currently being forced. When forcing a |Lazy| tagged lazy value, the
tag is first updated to |Forcing| using a |CAS|. Then, the first field is
updated to store the identifier of the current domain. Following this, the
deferred computation is evaluated.

If the deferred computation encounters a lazy value with |Forcing| tag and the
domain identifier in the first field matches the current domain identifier,
then it indicates recursive forcing of this lazy value. If the domain
identifiers are different, then it indicates concurrent forcing of this lazy by
more than one domain. In both of these cases, we raise the |Undefined|
exception.

\begin{wrapfigure}{r}{0.5\textwidth}
	\vspace{-0.4cm}
	\begin{code}[language=caml]
let rec safe_force l =
	match Lazy.try_force l with
	| None ->
		(* emits PAUSE instruction on x86 *)
		Domain.Sync.cpu_relax ();
		safe_force l
	| Some v -> v
	\end{code}
	\caption{Safely performing memoized computations in a concurrent setting}
	\label{fig:safe_force}
\end{wrapfigure}
If the computation results in a value, the first field is updated with the
result, and the tag is now updated to |Forward| using a |CAS|. If the
computation results in an exception, the first field is updated to a closure
which when evaluated raises the same exception. Then, the tag is reset to
|Lazy| using a |CAS| so that subsequent forcing of this lazy value raises the
|Undefined| exception. The GC also marks |Lazy| and |Forcing| tagged objects
with a |CAS| to handle the race between the mutator and the GC. Multicore-safe
lazy still uses the same object layout as stock OCaml, but uses two |CAS|es in
the mutator and one |CAS| for GC marking each lazy value.

This design permits the common case of using lazy values for memoization in a
concurrent setting. Given that we can distinguish between concurrent and
recursive forcing of a lazy value, Multicore OCaml provides a
\ml{Lazy.try_force: 'a Lazy.t -> 'a option} primitive which behaves similar to
\ml{Some (Lazy.force l)} except that it returns immediately with \ml{None} if
the lazy value \ml{l} is already being forced concurrently by another domain.
Using \ml{Lazy.try_force}, one may safely access memoized computations by
busy-waiting, for example, as shown in Figure~\ref{fig:safe_force}.

\subsection{Fibers}

Multicore OCaml supports lightweight concurrency through language-level threads
implemented using runtime support for heap-allocated, dynamically resized,
stack segments (fibers)~\cite{Dolan18}. While the language support for
concurrency is beyond the scope of the paper, care has to be taken towards the
interaction between the concurrent GC and fibers. Multicore OCaml uses a
deletion barrier which requires that the program stack be scanned at the start
of the cycle. Since fibers are program stacks, before switching control to a
fiber, all the objects on the fiber stack must be marked. This can lead to a
race between a mutator wanting to switch to a fiber and other domains marking
the fiber stack. Whenever a fiber stack is being scanned for marking, we obtain
a lock on the fiber. If a mutator attempts to switch to a locked fiber, it spin
waits until marking is complete. If a GC thread attempts to mark a locked
fiber, it skips marking since the fiber is being marked.

\section{Evaluation}
\label{sec:eval}

In this section, we evaluate the performance of the two Multicore OCaml GC
variants against the requirements listed in Section~\ref{sec:req}. We have
implemented the multicore support by extending the OCaml 4.06.1 compiler.

The performance evaluation was performed on a 2-socket, Intel\textregistered
Xeon\textregistered Gold 5120 x86-64 server, with 28 physical cores (14 cores
on each socket), and 2 hardware threads per core. Each core runs at 2.20GHz and
has 32 KB of L1 data cache, 32 KB of L1 instruction cache and 1MB of L2 cache.
The cores on a socket share a 19.25 MB L3 cache. The server has 64GB of main
memory and runs Ubuntu 18.04.

Since pause time measurements are sensitive to the operating system scheduling,
we carefully set up the machine to eliminate measurement noise. We disable
hardware threads, powersave and force a constant frequency of 2.20GHz. We
isolate 12 physical cores on each socket using Linux \texttt{isolcpus} for
benchmarking. The kernel will not schedule any processes on these isolated
cores unless explicitly requested. Thus, we have 12 cores each on 2 NUMA
domains for benchmarking.

\subsection{Sequential Performance}

In this section, we analyse the performance of sequential OCaml programs on
OCaml 4.06.1 (|Stock|), Multicore OCaml with the concurrent minor GC
(|ConcMinor|) and the stop-the-world parallel minor GC (|ParMinor|). We ran
|Stock| with default GC settings (next-fit collector, compaction is enabled).

Our aim is to determine whether unmodified sequential programs (modulo the
changes required for the C API on the |ConcMinor| variant) running on the
multicore variants retain throughput (running time) and latency (GC pause
times). The sequential benchmark suite comprises of a mix of workloads
including parsers (|menhir|, |setrip|, |yojson|), utilities (|cpdf|, |decompress|),
option pricing (|lexifi-g2pp|), ray tracer (|minilight|), concurrency
(|*lwt*|), bioinformatics (|fasta|, |knucleotide|, |revcomp2|, |regexredux2|),
numerical analysis (|matrix_multiplication|, |LU_decomposition|), and
simulation (|game_of_life|, |nbody|). The programs were run one after the other
on one of the isolated cores with no other load on the machine.

\subsubsection{Throughput}

\begin{figure}
  \begin{subfigure}{\textwidth}
		\includegraphics[scale=0.45]{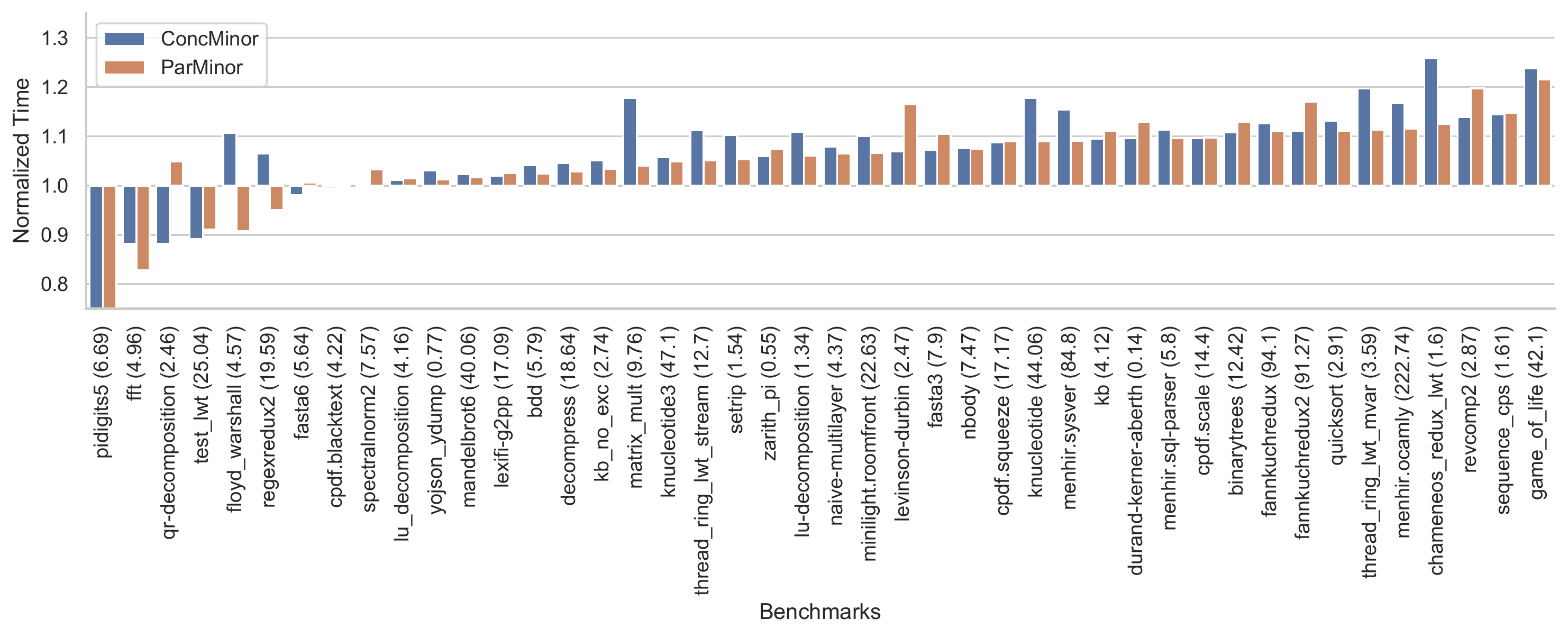}
		\caption{Normalized runtime. Baseline is \cf{Stock} OCaml whose
			running time in seconds is given in parenthesis.}
		\label{fig:seq_time}
	\end{subfigure}
	\begin{subfigure}{\textwidth}
		\includegraphics[scale=0.45]{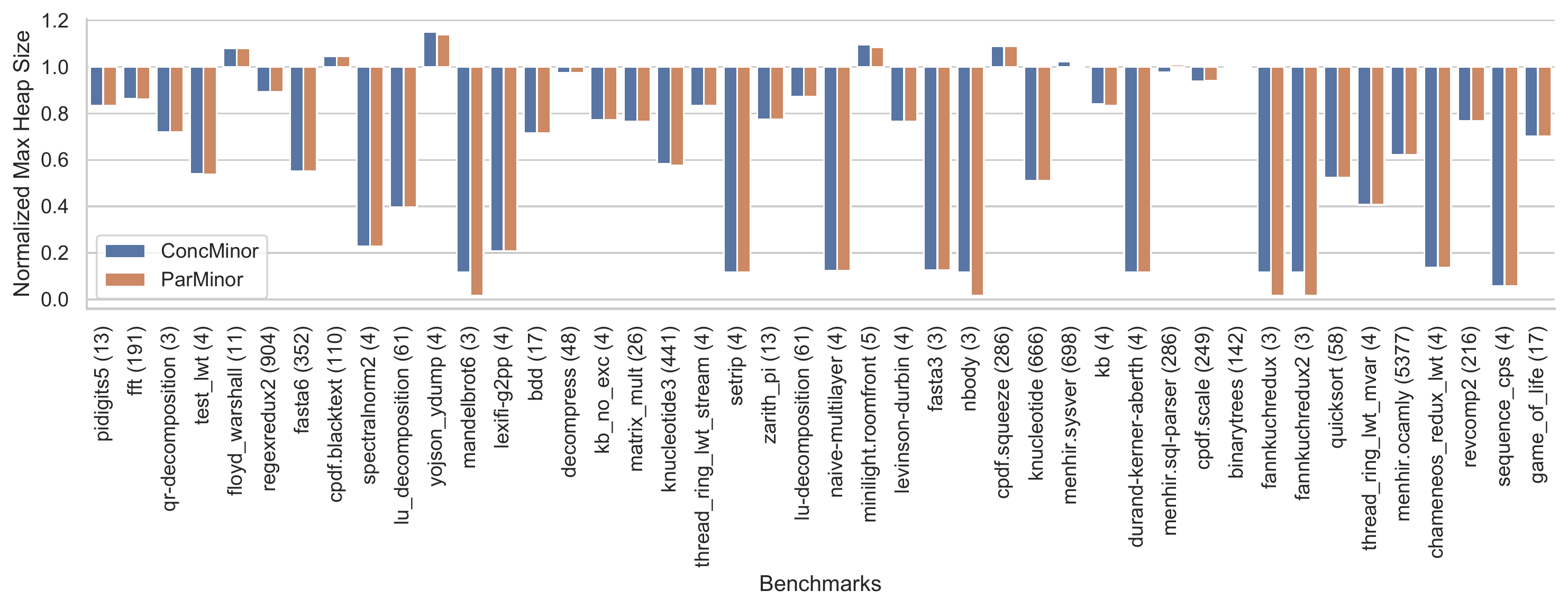}
		\caption{Normalized maximum major heap size. Baseline is \cf{Stock} OCaml
		whose maximum major heap size in MB is given in parenthesis.}
		\label{fig:seq_max_heap_size}
	\end{subfigure}
	\begin{subfigure}{\textwidth}
		\includegraphics[scale=0.45]{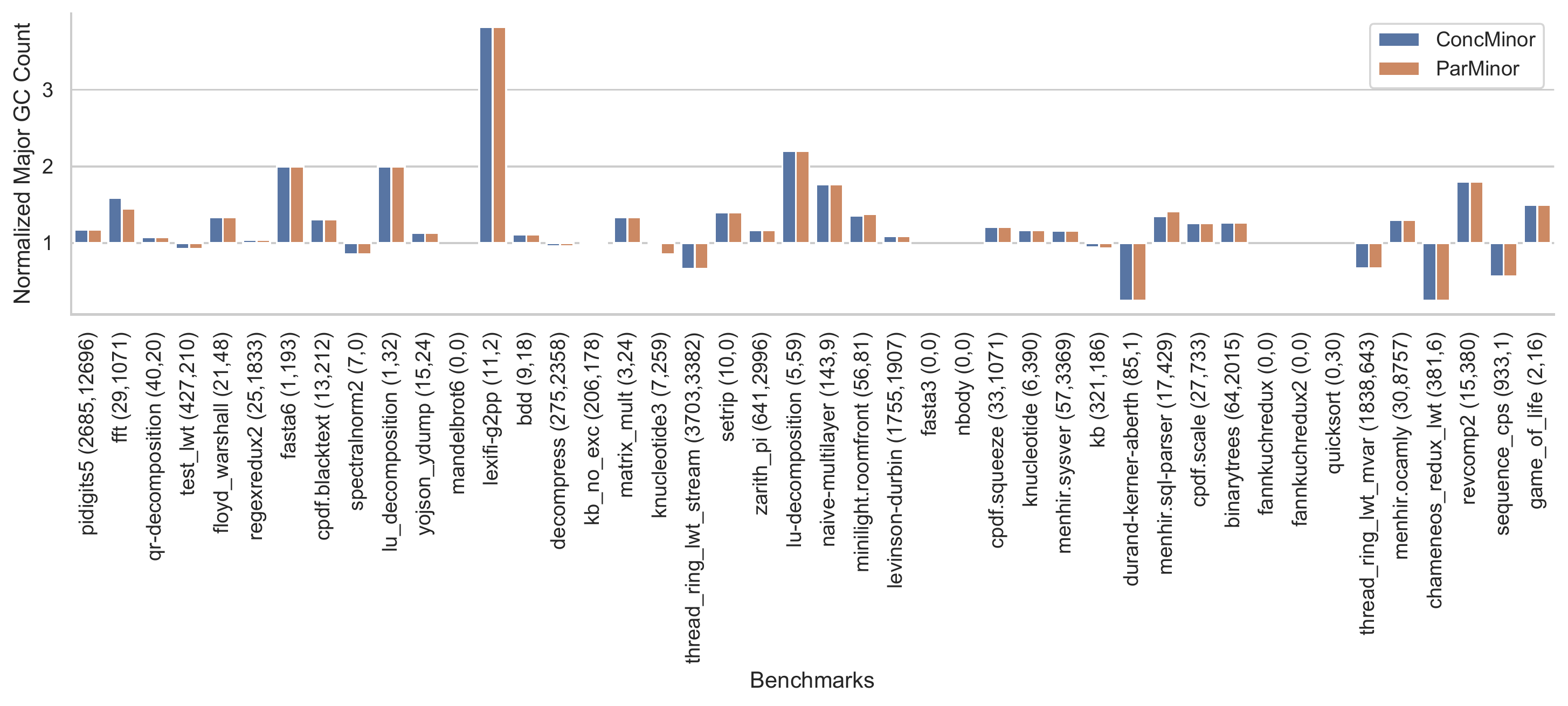}
		\caption{Normalized major collection count. Baseline is \cf{Stock} OCaml
		whose major collection count is given in first component in the
		parenthesis. The second component is the total allocation in the major heap
		in MB for \cf{Stock} OCaml. }
		\label{fig:seq_majgc_count}
	\end{subfigure}
	\caption{Throughput of unmodified sequential OCaml programs on Multicore OCaml}
\end{figure}

Figures~\ref{fig:seq_time},~\ref{fig:seq_max_heap_size}
and~\ref{fig:seq_majgc_count} respectively show the normalized running time,
the normalized maximum major heap size and the normalized major collection
count of the sequential benchmarks on Multicore OCaml with |Stock| OCaml
running time as the baseline.

On average (using geometric mean of the normalized running times), |ConcMinor|
and |ParMinor| are 4.9\% and 3.5\% slower than |Stock|, respectively.
|ConcMinor| is slower than |ParMinor| due to the overhead of the read barrier.
This overhead itself manifests in two ways. First is the additional
instructions that |ConcMinor| will have to execute over |ParMinor|. Secondly,
since the read barrier is inlined at reads, the additional instructions tend to
affect instruction cache efficiency. We have also observed that additional
instructions due to read barrier in tight loops sometimes tend to disable or
enable micro architectural optimisations, which tend to affect overall running
time by a few percentage points. Hence, we believe that observed overheads are
within reasonable bounds.

Focussing on specific benchmarks, |pidigits5| on |Stock| is 2.9$\times$ slower than
the both of the multicore variants. This is due to the fact that |pidigits5|
allocates 12.7GB in the major heap in total, most of which are short-lived.
This causes |Stock| to schedule 895 compactions and the benchmark spends 50\%
of its running time in the kernel allocating and freeing memory. By turning off
compaction for |pidigits|, the runtime reduces to 2.21 seconds which is
5\%(9\%) faster than |ConcMinor|(|ParMinor|).

On the other side, |game_of_life| on multicore variants is 20\% slower than
|Stock|. |game_of_life| uses exceptions to encode the behaviour on
out-of-bounds array access. Exceptions in the multicore variants are more
expensive than |Stock| for two reasons. Firstly, the dedicated exception
pointer register on x86-64 is now used for the pointer to the domain-local
runtime structure, which holds the domain-local state including the exception
pointer. Secondly, due to fibers in the multicore runtime, the stacks need to
move to grow on stack overflow, unlike |Stock|. While |Stock| uses absolute
addresses into the stack for exception pointers, we use relative addressing to
encode the exception pointer to avoid rewriting the exception pointer when
stacks are moved. Hence, the multicore variants need to execute additional
instructions for setting up the exception handler and raising exceptions. We
verified this by rewriting the benchmark using an optional type to capture the
exception behaviour and the performance matches that of |Stock|. We
experimented with the idea of retaining the dedicated exception pointer
register in addition to the register for domain-local runtime state. In this
case we observed performance that was generally worse; there was increased
register pressure and consequently more spilling. Implementing absolute
addressing for exceptions in the multicore variants would improve performance
of exceptions, while making the (rarer) stack move operations slower.

On average, |ConcMinor| and |ParMinor| use 54\% and 61\% less memory than
|Stock| (Figure~\ref{fig:seq_max_heap_size}). Many of the benchmarks where the
multicore variants do better than |Stock| do not use much memory. However,
multicore variants do better even on those benchmarks that allocates a lot of
memory. For example, |menhir.ocamly| uses a maximum heap size of 5.3 GB on
|Stock| and the multicore variants consume 3.3 GB. The next-fit collector that
we use in |Stock| is susceptible to fragmentation whereas the size-segmented
free list used in the multicore variants are tolerant to fragmentation. We ran
the benchmark with first-fit collector and the benchmark did not terminate
after 30 minutes. With the new best-fit collector which was added in OCaml
4.10.0, the benchmark runs in 250 seconds and consumes 3.3 GB, which matches
the behaviour of multicore variants. The results show that the runtime and
memory usage of sequential OCaml programs on the multicore variants is
competitive with stock OCaml.

In our experiments, we use the same minor heap size (the OCaml default of 2 MB)
for all the three variants. We observed that all the three variants allocate
similar amount memory in the major heap. The average difference was less than
1\%. Hence, we report only the total allocations in the major heap for |Stock|
(Figure~\ref{fig:seq_majgc_count}). We observe that the multicore variants tend
to do a few more major collections than |Stock|. Although the variants allocate
similar amounts of memory in the major heap, the allocators are different
between |Stock| and the multicore variants, and vary in the amount of
fragmentation in the heap. This induces differences in the GC major slice
scheduling algorithm causing more major collections in the multicore variants.

\subsubsection{Latency}

\begin{figure}
  \begin{subfigure}{\textwidth}
		\includegraphics[scale=0.47]{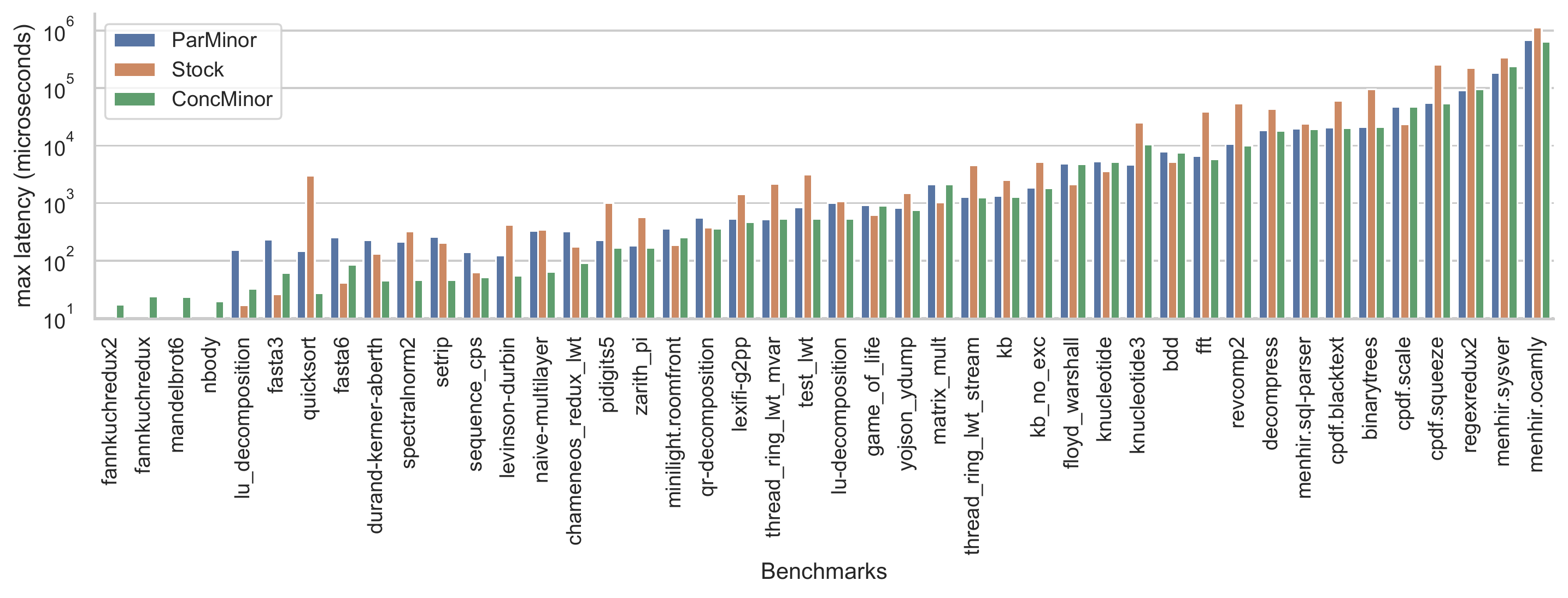}
		\caption{Max pause times.}
		\label{fig:max_latency}
	\end{subfigure}
	\begin{subfigure}{\textwidth}
		\includegraphics[scale=0.47]{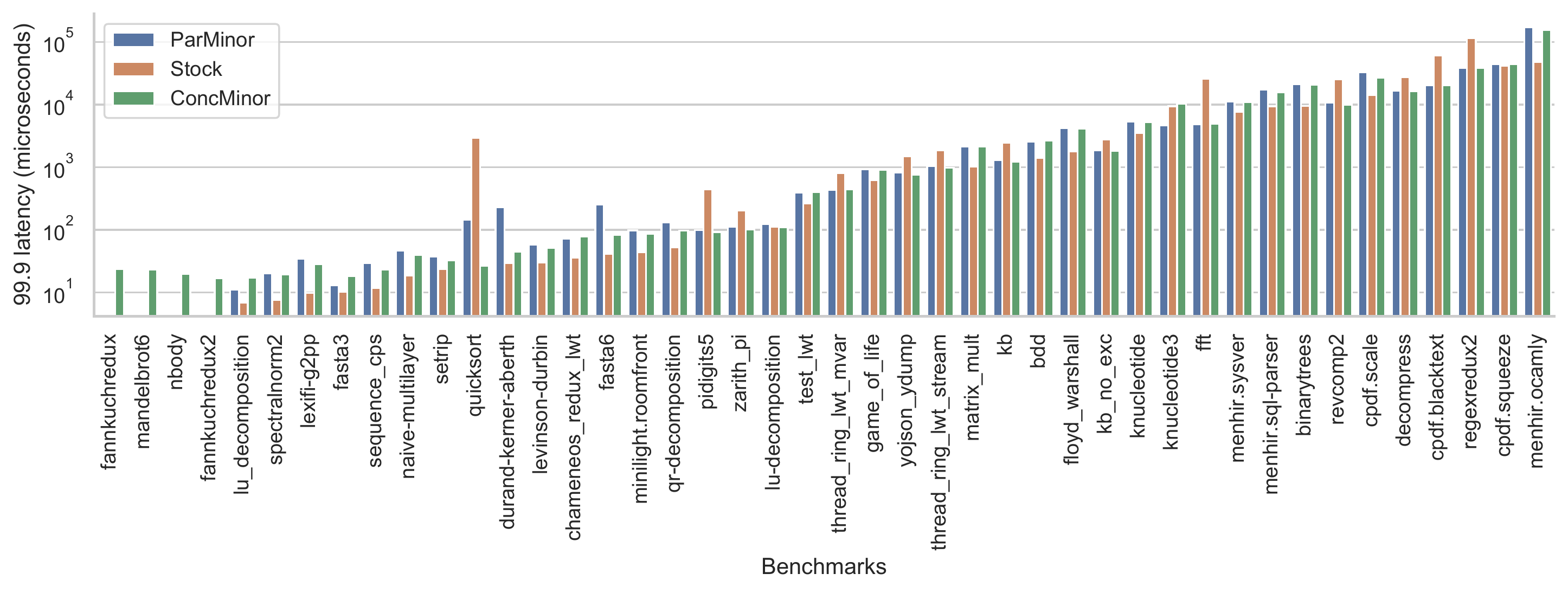}
		\caption{99.9th percentile pause times.}
		\label{fig:999_latency}
	\end{subfigure}
  \caption{Pause times of unmodified sequential OCaml programs}
\end{figure}
\vspace{-0.5cm}

Figure~\ref{fig:max_latency} and~\ref{fig:999_latency} shows the maximum and
99.9th percentile GC pause times of unmodified OCaml programs running on the
different variants. We observe that both of the multicore variants remain
competitive with |Stock|.

The most unresponsive benchmark is |menhir.ocamly| where the maximum pause times
are 1125 ms, 689 ms and 635 ms on |Stock|, |ParMinor| (39\% lower than |Stock|)
and |ConcMinor| (44\% lower then |Stock|). There are a number of benchmarks
whose maximum pause times are over 10 ms. We observed that in all of these
programs, the maximum pause time is attributed to a large slice of major GC
work.

While the GCs are incremental, the runtime computes a \emph{mark slice budget},
the amount of incremental work (in memory words) that should be done in each
cycle. The budget is computed based on the \emph{space overhead} setting
(percentage of memory used for live data), the current size of the major heap,
the allocations since the last major slice, and a summarised history of work
done in previous slices. In these benchmarks, there is a large amount of
allocation since the last slice of major work, which causes the runtime to
compute a very large budget, which in turn causes the large pause times.
However, such huge imbalances are rare events. We can observe that the 99.9th
percentile pause times are much better than the maximum pause times. For
|menhir.ocamly|, 99.9th percentile pause times for |Stock|, |ConcMinor| and
|ParMinor| are 47 ms, 155 ms and 171 ms, respectively.

While these benchmarks are not interactive programs, the large pause times
observed are antithetical to the common wisdom that OCaml is suitable for
interactive programs. Are these large pause times indicative of the fundamental
lack of responsiveness in the GC design or are they an artefact of the GC
pacing algorithm? OCaml does not have an explicit setting for responsiveness;
the space overhead parameter controls memory usage and only indirectly affects
the pause times. We first observed that increasing the space overhead did not
bring down the maximum pause time in |menhir.ocamly|.

We experimented with a threshold for the maximum budget for a mark slice. The
threshold was chosen such that no major slice is longer than roughly 10 ms on
the benchmarking machine. Without the maximum budget, |menhir.ocamly| on
|ConcMinor| completed in 213 s, with a maximum major heap size of 3.3 GB and a
maximum latency of 635 ms. With the maximum budget, the benchmark completed in
203 s, with a maximum major heap size of 6.07 GB and a maximum latency of 10
ms. While this approach will not ensure that the space overhead constraints are
met, we are able to trade space for improvements in both the running time as
well as maximum latency. This illustrates that our GC design does not
fundamentally lack responsiveness.

\subsection{Parallel Benchmarks}

With parallel benchmarking, our aim is to determine whether the multicore
variants scale with additional cores, and how the GC pause times are affected
by additional core count. Our parallel benchmark suite is composed of numerical
kernels (|LU_decomposition|, |matrix_multiplication|, |spectralnorm2|), ray
tracing (|minilight|), GC stress test mirroring real workloads (|binarytrees5|),
Zlib compression and decompression (|test_decompress|), simulations
(|game_of_life|, |nbody|), all-pairs shortest path algorithm (|floyd_warshall|)
and Mandelbrot set plotting (|mandelbrot6|).

As mentioned earlier, we had isolated 12 cores each on the two sockets in the
server. We ran each benchmark with the same input at different domain counts.
For the domain counts less than or equal to 12, we pin them to the same socket
using Linux \texttt{taskset} command. By default, Linux allocates pages in the
same NUMA domain as the process generating the request. Hence, pinning the
domains to the same NUMA domain minimises NUMA effects. For domain counts
greater than 12, they are allowed to span both sockets.

\begin{figure}
	\includegraphics[width=\textwidth]{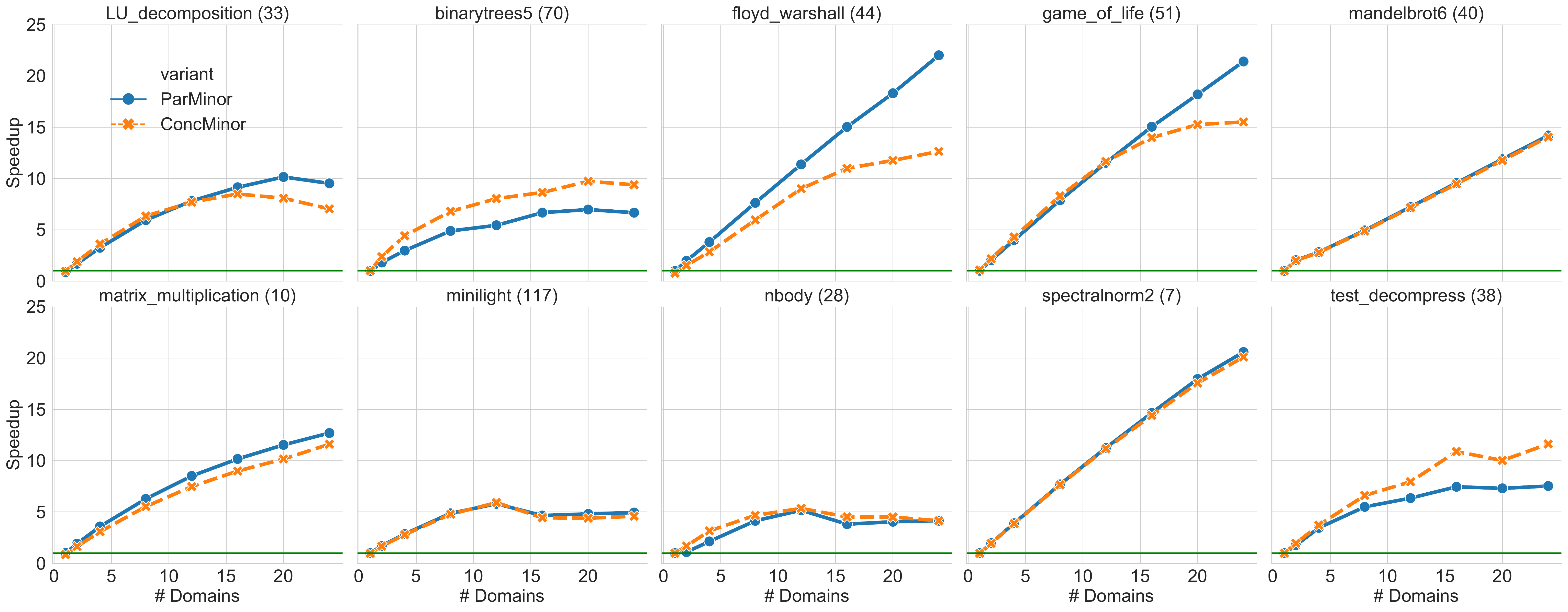}
	\caption{Speedup of parallel benchmarks. The baseline running time in seconds
	is given in the parenthesis next to the benchmark name. The line parallel to
	the x-axis is at y=1.}
  \label{fig:parallel_time}
\end{figure}

Figure~\ref{fig:parallel_time} shows the speedup of the multicore variants. The
baseline for speedup is the fastest sequential version of the program. The
results show that |ParMinor| scales linearly for |floyd_warshall|,
|game_of_life| and |spectral_norm|, whereas |ConcMinor| scales linearly only
for |spectral_norm|. All of these benchmarks have a number of iterations with
parallelisable work available in each iteration. In each iteration, the master
domain farms out the work to the worker domains over a point to point channel
data structure and synchronizes with the workers at the end.

In the case of |ConcMinor|, since the channels themselves are shared between
the domains, they are in the major heap. When the master adds work in the
channel, a pointer is created from the channel in the major heap to the work in
the master domain's minor heap arena. When the worker fetches the work, it
faults on the read and has to send an interrupt to the master domain to promote
the work to the major heap. Until the master finishes promoting the work, the
worker sits idle. This causes the gradual tapering out of speedup with
increasing domain count. The master can eagerly promote the work, but this too
has non-trivial overheads since at least the roots have to be scanned to ensure
that there are no dangling references to the promoted object
(Section~\ref{sec:promotion}). On the other hand, in the case of |ParMinor|,
the worker directly reads the work item from master's minor heap arena. The
other benchmarks have large non-parallelisable sections which causes sublinear
speedup. Overall, when there is parallelisable work available, Multicore OCaml
can effectively parallelise it.

\begin{table}
	\begin{tabular}{lrrrrrrrr} \toprule
		\multirow{2}{*}{Benchmark} & \multicolumn{2}{c}{Major GCs} & \multicolumn{2}{c}{Minor GCs} & \multicolumn{2}{c}{Major Allocs (MB)} & \multicolumn{2}{c}{Max Heap (MB)} \\ \cmidrule(r){2-3} \cmidrule(r){4-5} \cmidrule(r){6-7} \cmidrule{8-9}
															 & Conc &       Par & Conc &       Par & Conc               & Par & Conc &                Par \\ \midrule
		|LU_decomposition|         & 98 & 15 & 66402 & 83469 & 141 & 134 & 100 & 101 \\
		|binarytrees5|             & 20 & 17 & 14084 & 17800 & 8088 & 7729 & 5124 & 7752 \\
		|floyd_warshall|           & 15 & 15 & 409 & 424 & 19 & 13 & 13 & 13 \\
		|game_of_life|             & 3 & 3 & 2 & 9 & 12 & 12 & 12 & 12 \\
    |mandelbrot6|              & 0 & 0 & 0 & 0 & 1 & 1 & 1 & 1 \\
		|matrix_multiplication|    & 0 & 0 & 0 & 0 & 0 & 0 & 24 & 24 \\
		|minilight|                & 44 & 6 & 116433 & 210146 & 272 & 249 & 48 & 166 \\
		|nbody|                    & 1 & 1 & 1 & 2 & 0 & 0 & 0 & 0 \\
		|spectralnorm2|            & 12 & 6 & 8249 & 10152 & 1 & 1 & 3 & 5 \\
		|test_decompress|          & 16 & 4 & 22177 & 24218 & 5157 & 3693 & 2239 & 3851 \\
		\bottomrule
  \end{tabular}
	\caption{GC statistics for parallel benchmarks with 24 domains.}
	\label{tab:stat24}
\end{table}

Table~\ref{tab:stat24} presents the GC statistics for the benchmark runs with
24 domains. We observe that |ParMinor| performs more minor GCs than
|ConcMinor|. In the case of |ParMinor| whenever the heap arena of a domain
fills up, the domain forces all the other domains to also collect their minor
heaps. This issue can be fixed by having the domains request large pages for
allocations from a contiguous minor heap space such that the minor collection
is only performed when all of the minor heap is full. We also observe that
|ConcMinor| allocate more in the major heap. This is due to the fact that any
work that is shared between the domains need to be promoted to the major heap,
whereas |ParMinor| can share without promotion.

\begin{figure}
	\includegraphics[width=\textwidth]{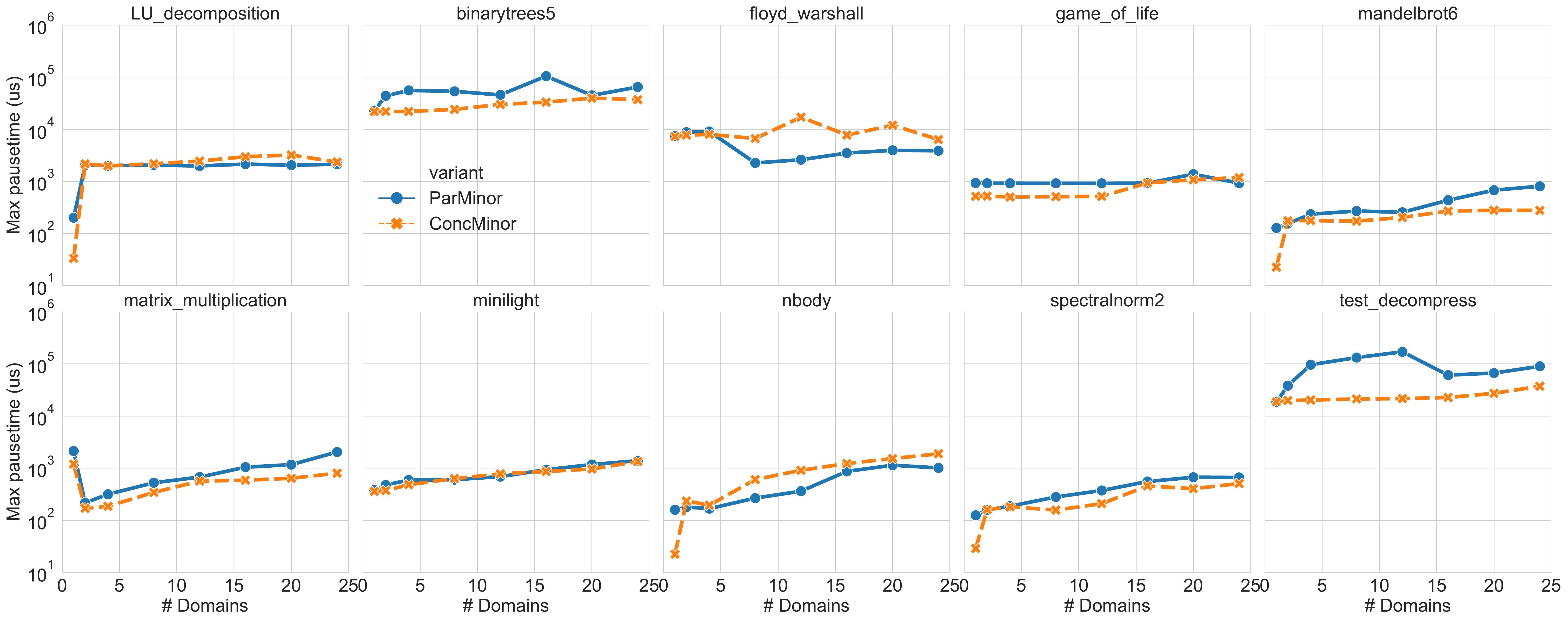}
	\caption{Max GC pause times of parallel benchmarks.}
  \label{fig:parallel_pause}
\end{figure}

Figure~\ref{fig:parallel_pause} shows the maximum GC pause times for the parallel
benchmarks with increasing number of domains. We observe that the pause times
of both the collectors are tolerant to increase in the number of domains. The
maximum pause times remains the same or increases only marginally with
increasing domain count. The worst case pause times for |ParMinor| is on
|test_decompress| (170 ms at 12 domains) and |ConcMinor| is on |binarytrees5|
(76 ms at 20 domains). For comparison, the maximum pause time for the
sequential versions of the benchmarks on |Stock| OCaml are 43 ms for
|test_decompress| and 94 ms for |binarytrees5|. Hence, the observed maximum
latency numbers are within acceptable bounds.

The maximum pause times were observed when a domain finishes a minor cycle and
then runs a slice of the major heap work where it broadcasts a request for a
stop-the-world barrier. An immediate improvement would be to separate out minor
collection from the major slice work, as is the case in |Stock|, so that they
are performed in two separate pauses. Hence, there is room for further
improvements in pause times.

\section{Discussion}
\label{sec:discussion}

As we noted at the beginning of this paper, bringing parallelism to OCaml
required us to find a point in the GC design space where sequential single core
performance can be preserved but still introduce scalable parallelism as an
option to OCaml developers. In this section, we will discuss the difficult
design decisions that relate to this task of retrofitting parallelism onto an
existing venerable functional language with a large existing codebase.

The core OCaml language had to be constrained in surprisingly few places where
it exposed too much of the single-core internals. The |Obj| module from the
standard library is the biggest culprit; even though it is advertised as ``not
for the casual user'', it is often used in the result of extracted code (e.g.
from Coq) and the compiler itself. We primarily only had to remove two
operations which update the header field of an object -- |Obj.set_tag| and
|Obj.truncate|. These operations conflict with concurrent GC threads
non-atomically marking the object. Alternative primitives are provided that
expose higher level functionality for some of the common use cases found.

Most of the changes elsewhere happened towards the parts of the language
involving interfacing with the outside world. Signal handling, parts of the
FFI, and marshalling required some modifications to make their semantics more
robust in the face of parallelism, primarily by removing race conditions or
shared state. In most cases, we simply had to tighten up the definition of
allowable behaviour: for example, named callbacks to OCaml from C can no longer
be mutated once registered using the raw pointer returned. We found no cases of
this causing a problem in real world code.

The more advanced language features that had to be adapted for Multicore OCaml
(ephemerons, lazy values and finalisers) are designed to minimise single-core
performance impact. For example, lazy values introduce just two |CAS|
operations into the value-forcing path, but none if already forced. Given that
lazy values are introduced to memoize expensive computations the cost of the
(almost always uncontended) |CAS| gets amortized. Multicore OCaml does not
support compactions. We observed in our experiments that the size-segmented
multicore allocator is more tolerant to fragmentation and performs as well as
the |best-fit| collector added in OCaml 4.10.0 on |menhir.ocamly|. The result
of this careful design is that OCaml users should be able to move gradually
over to adopting multicore parallelism, without finding regressions (either
semantically or performance-wise) in their existing large codebases.

When assessing the performance impact of our designs, we had to construct a
micro- and macro-benchmark suite that is assembled from the wider OCaml
community (primarily by extracting representative samples from the |opam|
package manager).  We immediately observed the difficulty of even small
backwards compatibility changes between stock and multicore OCaml being
difficult to adapt to in our own evaluations.

One primary culprit was the C API changes required by our original concurrent
minor collector, which motivated us to build alternative designs to firmly
determine the cost of adding read barriers. As we showed in our evaluation
(\S\ref{sec:eval}), we found that our stop-the-world parallel minor collector
that did not require changes to the OCaml FFI maintained extremely low pause
times and good throughput.  This happy result means that the multicore OCaml
retrofit can be integrated into the mainstream OCaml distribution with very few
user-observable compatibility concerns.

Multicore OCaml has also chosen to separate parallelism at the runtime level
from concurrency at the language level.  There is active research ongoing into
various ways to integrate concurrency into OCaml via algebraic
effects~\cite{dolan18effects, kc18effects}.  From the perspective of the
runtime, we have exposed just enough functionality via lightweight fibers to
allow this research into strongly typed concurrent OCaml extensions to
continue, but also not block the broader OCaml community from directly using
the parallelism by spinning up domains using relatively low-level interfaces.

\section{Related Work}
\label{sec:related}

There are several notable works in the recent past that extend languages in the
ML family with support for parallel programming. Manticore~\cite{Fluet10,
Auhagen11} is a high-level parallel programming language with support for
CML~\cite{ReppyCML}-style coarse-grained parallelism as well as fine-grained
parallelism in the form of parallel tuple expressions and parallel array
comprehensions. The core language is a subset of Standard ML without mutable
reference cells and arrays. MaPLe~\cite{Raghunathan16, Guatto18, Westrick20} is
a parallel extension of the MLton Standard ML compiler with support for nested
parallelism. The programming model, as well as the garbage collector in MaPLe,
takes advantage of \emph{disentaglement} property which mandates that a thread
cannot witness the memory allocations of another concurrent thread. Unlike
these approaches, Multicore OCaml places no restriction on the programming
model and aims to be fully compatible with the stock OCaml programming
language. Multicore OCaml also places no restrictions on the object graph that
can be constructed and shared between multiple domains.

SML\# features a concurrent, non-moving collector~\cite{ueno16gc} that is a
similar approach to Multicore OCaml, but a single heap instead of a
generational one. Multicore OCaml retains a generational approach in order to
reduce the pressure on the major heap, at the cost of more complexity in the
interaction between generations. SML\# allocator uses BiBoP~\cite{Steele77}
technique to distinguish used and free blocks in the free list to avoid adding
a header to each object. Since OCaml objects already include a header, we reuse
it in the free list and avoid having to use an auxiliary data structure.
\cite{ueno16gc} collector only supports a single GC thread. The paper also does
not mention whether advanced data structures such as ephemerons, weak
references and finalisers are supported. We observed that with multiple GC
threads, careful attention has to be paid to the design of advanced data
structures.

GHC has recently integrated an incremental, non-moving collector also
inspired by the design in SML\#~\cite{gamari20}. GHC also maintains a
nursery heap to allocate into rather than directly going into the major heap.
Similar to SML\#, only one GC thread is currently supported with ongoing work
to support multiple threads.

We retain the option to use the concurrent minor collector in a future revision
of Multicore OCaml if it turns out to be necessary for many-core scalability.
Although our stop-the-world minor collector scaled up admirably to at least 24
cores, synchronisation costs will inevitably increase as the number of cores
do.  Adding a read barrier (as used by the concurrent minor collector) opens up
the GC design space considerably; for example to add a pauseless
algorithm~\cite{cliff05pauseless} such as the ones used in modern Java GCs like
Azul, ZGC~\cite{tene11pauseless} or Shenandoah~\cite{flood16shen}.  Our results
for the concurrent minor collector show that the read barrier has less overhead
in OCaml code than initially expected, and would be a reasonable solution in
many-core machines.

Several previous works~\cite{Doligez93, Auhagen11, Anderson10, Domani02,
Marlow11, Sivaramakrishnan14} have explored concurrent minor collection with
thread-private minor heap arenas. The \cite{Marlow11} collector for GHC allows
major to minor pointers and on-demand promotion, similarly to us. Our
innovation here is the use of virtual memory mapping techniques in our
concurrent minor collector (\S\ref{sec:readbarrier}) to optimise the read
barrier required to trap reads to a foreign minor heap arena.

Manticore~\cite{Auhagen11} and MultiMLton~\cite{Sivaramakrishnan14} use
Appel-style semi-generational collection for the threads-local heaps. The local
heap supports both minor and major collections. There is a dedicated shared
heap in order to share objects between multiple threads. Similar to us
Manticore permits pointers from the shared heap to the local heaps, and
on-demand promotion. MultiMLton on the other hand does not permit pointers from
the shared heap to the local heap, but takes advantage of ample user-level
concurrency to preempts the execution of user-level threads that are about to
introduce a shared to local heap pointer. In Manticore, the global heap is
collector with a stop-the-world parallel collector, while MultiMLton uses a
stop-the-world serial collector. In order to retain the low pause times of
OCaml, Multicore OCaml uses concurrent mark-and-sweep collector for the major
(shared) heap.

The Go programming language uses a concurrent, tri-colour, non-moving,
mark-and-sweep, generational garbage collector and is designed for low GC
pausetimes. Of particular interest is the interaction between goroutines and
the GC. Before Multicore OCaml switches to an unmarked fiber, all the objects
on the fiber stack are marked. This is necessary since Multicore OCaml uses
a deletion barrier. Go opts for a different design where the goroutines are not
marked before switching control to them. In turn, Go's write barrier is a
combination of deletion and insertion barrier; until the current goroutine is
shaded black, the write barrier marks both the object overwritten and the
referrent~\cite{GoWriteBarrier}. This design will be more responsive in
concurrency heavy programs, at the cost of making write barriers more expensive
even for code that does not use lightweight concurrency.

Our overall problem of retrofitting parallelism onto an existing sequential
language has previously been explored by the Racket programming language as
well.  They added parallelism to a sequential runtime via two separate methods;
firstly via a |futures| library that classifies operations that are
parallel-safe as separate from ones that required more
synchronisation~\cite{10.1145/1869459.1869507}, and secondly via a |places|
library that provides message-passing parallelism~\cite{10.1145/2047849.2047860}.
This approach is significantly less invasive to the Racket runtime than Multicore
OCaml is to current mainline OCaml, but the parallel performance is reported
as best suited to numerical tasks.  Multicore OCaml aims to provide good
parallel performance for algorithms involving large shared data structures,
(such as those found in symbolic processing domains such as proof assistants),
and also to integrate with existing debugging and tracing tools. Thus, we
judged the significant engineering effort required to build the Multicore OCaml
garbage collector as appropriate.

\section{Conclusions}
\label{sec:conc}

We present a novel, mostly-concurrent garbage collector that has been designed
with the explicit goal of retrofitting parallelism onto OCaml while preserving
performance and feature backwards compatibility. The extensive experimental
evaluation validates that the new collector not only preserves backwards
compatibility but also scale admirably on multiple cores.

All of the artefacts emerging from this research will be made available as
open-source software and patches to the upstream OCaml distribution. The
development repositories are available at
\url{https://github.com/ocaml-bench/sandmark} for the benchmarking suite used
in this paper, \url{https://github.com/ocaml-multicore/ocaml-multicore} for the
compiler forks, and
\url{https://github.com/ocaml-multicore/multicore-ocaml-verify} for the SPIN
models.

\section*{Acknowledgements}

We thank the anonymous reviewers, Fran\c cois Pottier, Pierre Chambart, Jon
Harrop, Josh Berdine, Sam Goldman, and Guillaume Munch-Maccagnoni for their
feedback on earlier drafts of the paper, Xavier Leroy and Damien Doligez and
the core OCaml development team for their comments on the design of the
Multicore OCaml GC. Portions of this research was funded via a Royal Commission
for the Exhibition of 1851 and Darwin College Research Fellowships, and by
grants from Jane Street and the Tezos Foundation.

\newpage
\bibliography{retro-parallel}

\ifdefined\techreport
\newpage
\appendix
\section{Major slice with ephemerons and finalisers}
\label{sec:slice3}

\begin{code}
/* Assume: pushing an object into an empty mark stack will increment
	[gNumDomsToMark] and set [dlMarkingDone = false]. */
\end{code}
\begin{minipage}[t]{0.47\textwidth}
\begin{code}[numbers=left]
def majorSlice (budget):
	budget = sweepSlice (budget)
	budget = markSlice (budget)
	if (budget && !dlMarkingDone):
		dlMarkingDone = 1
		atomic:
			gNumDomsToMark--
			gEpheRound++
			gNumDomsMarkedEphe = 0

	/* Finalisers */
	if (Phase == MARK_FINAL):
		mark_final ()
		gNumDomsMarkedFinal++
	if (Phase == SWEEP_EPHE):
		mark_final_last ()
		gNumDomsMarkedFinalLast++

	/* Ephemeron Mark */
	cached = gEpheRound
	if (gPhase == MARK_FINAL &&
	    cached > dlEpheRound):
		budget=markEphe(budget,cached)
		if (budget && dlMarkingDone)
			dlEpheRound = cached
			atomic:
				if (cached == gEpheRound):
					gNumDomsMarkedEphe++

	/* Ephemeron Sweep */
	if (gPhase == SWEEP_EPHE):
		budget = sweepEphe(budget)
		if (budget && !dlSweepEpheDone):
		  dlSweepEpheDone = 1
			atomic: gNumDomsSweptEphe++

	/* Change Phase */
	changePhase ()
\end{code}
\end{minipage}
	\begin{minipage}[t]{0.51\textwidth}
\begin{code}[numbers=left,firstnumber=39]
def changePhase ():
	if (gPhase == MARK && gNumDomsToMark == 0):
		barrier:
			if (gPhase == MARK_FINAL &&
			    gNumDomsToMark == 0):
				gPhase = MARK_FINAL
	if (gPhase == MARK_FINAL &&
	    gNumDomsToMark == 0 &&
			gNumDomsMarkedEphe == gNumDoms &&
			gNumDomsMarkedFinal == gNumDoms):
		barrier:
			if (gPhase == MARK_FINAL &&
			    gNumDomsToMark == 0 &&
					gNumDomsMarkedEphe == gNumDoms &&
					gNumDomsMarkedFinal == gNumDoms):
				gPhase = SWEEP_EPHE
	if (gPhase == SWEEP_EPHE &&
	    gNumDomsSweptEphe == gNumDoms &&
			gNumDomsMarkedFinalLast == gNumDoms):
		runOnAllDoms (cycleHeap)

def cycleHeap():
	barrier:
		newc.Unmarked = gColours.Marked
		newc.Garbage = gColours.Unmarked
		newc.Marked = gColours.Garbage
		newc.Free = gColours.Free
		gColours = newc
		gNumDomsToMark = gNumDoms
		gEpheRound = gNumDomsMarkedEphe = 0
		gNumDomsSweptEphe = 0
		gNumDomsMarkedFinal = 0
		gNumDomsMarkedFinalLast = 0
		gPhase = MARK
	dlMarkingDone = dlEpheRound = 0
	dlSweepEpheDone = 0
	markNonIncrementalRoots ()
\end{code}
\end{minipage}
\fi

\end{document}